\documentclass[pra,aps,eqsecnum,showpacs,superscriptaddress]{revtex4}
\usepackage{multirow}  
\usepackage{amsmath}
\usepackage{amssymb}
\usepackage{latexsym}
\usepackage{float}
\usepackage{graphicx}
\usepackage{bm}

\makeatletter
\newcommand{\lbar}{{\mathchoice
  {\smash@bar\textfont\displaystyle{0.25}{1.2}\lambda}
  {\smash@bar\textfont\textstyle{0.25}{1.2}\lambda}
  {\smash@bar\scriptfont\scriptstyle{0.25}{1.2}\lambda}
  {\smash@bar\scriptscriptfont\scriptscriptstyle{0.25}{1.2}\lambda}
}}
\newcommand{\smash@bar}[4]{%
  \smash{\rlap{\raisebox{-#3\fontdimen5#10}{$\m@th#2\mkern#4mu\mathchar'26$}}}%
}
\makeatother

\begin{document}
\newcommand{\ket} [1] {\vert #1 \rangle}
\newcommand{\bra} [1] {\langle #1 \vert}
\newcommand{\braket}[2]{\langle #1 | #2 \rangle}
\newcommand{\proj}[1]{\ket{#1}\bra{#1}}
\newcommand{\mean}[1]{\langle #1 \rangle}
\newcommand{\opnorm}[1]{|\!|\!|#1|\!|\!|}
\newcommand{\hr}[1]{\hat{\rho}_{#1}}
\newcommand{\tr}[1]{{\rm Tr}\,\hr{#1}}
\title{Classical Light Beams and Geometric Phases}

\author{N. Mukunda}
\email{nmukunda@gmail.com} 
\affiliation{Optics \& Quantum Information Group, The Institute of Mathematical
  Sciences, C.I.T Campus, Tharamani, Chennai 600 113, India.}

\author{S. Chaturvedi}
\email{scsp@uohyd.ernet.in} 
\affiliation{School of Physics, University of Hyderabad, Hyderabad 500 046, India}

\author{R. Simon}
\email{simon@imsc.res.in} 
\affiliation{Optics \& Quantum Information Group, The Institute of Mathematical
  Sciences, C.I.T Campus, Tharamani, Chennai 600 113, India.}

\begin{abstract}
We present a study of geometric phases in classical wave and polarisation optics 
using the basic mathematical framework of quantum mechanics. 
Important physical situations taken from scalar wave optics, 
pure polarisation optics, and the behaviour of polarisation in the 
eikonal or ray limit of Maxwell's equations in a transparent medium 
are considered. The case of a beam of light whose propagation direction and 
polarisation state are both subject to change is dealt with, 
attention being paid to the validity of Maxwell's equations at all stages. 
Global topological aspects of the space of all propagation directions 
are discussed using elementary group theoretical ideas, 
and the effects on geometric phases are elucidated. \\

~~{\em OCIS codes}: (030.1640) Coherence; (350.5500) Propagation; 
  (260.5430) Polarization
\end{abstract}

\maketitle
\section{Introduction}
The quantum mechanical geometric phase was discovered by Berry in
1983\,--\,84\,\cite{berry}. The context was unitary evolution governed by the
Schr\"odinger equation in the adiabatic approximation, i.e., with a
hermitian Hamiltonian possessing a `gentle' time-dependence.
Assuming that as an operator the Hamiltonian is cyclic, i.e., it
returns to its original form after a certain interval of time
(during which there are no level crossings), the approximate solutions
to the Schr\"odinger equation are also cyclic. The geometric phase
is then seen explicitly in these solutions at the end of the cycle.

Immediately after Berry's discovery, it was pointed out by Barry 
Simon\,\cite{barry} that the geometric phase expresses the non-integrability, 
or anholonomy, of a natural `rule of parallel transport' (a connection) in a 
principal fibre bundle, with structure group $U(1)$, which occurs in the 
framework of quantum mechanics. This was therefore a characterisation of this 
phase in the language of differential geometry.

The ensuing years saw two streams of work relating to the geometric
phase. One consisted of extensions of Berry's original work, in the
sense of relaxing the conditions under which the phase is definable.
The other consisted of interesting earlier results which could be
reinterpreted as instances of this phase, and so as precursors to it. We
recall three significant efforts of the first kind. Aharonov and
Anandan\,\cite{anandan} showed that the adiabatic condition is not necessary---given 
a cyclic solution to the Schr\"odinger equation involving any
(time-dependent) Hamiltonian, one can reconstruct a corresponding
geometric phase. This was followed by the work of Samuel and
Bhandari\,\cite{samuel}, in which the cyclic condition on a solution was also
dispensed with. Given a solution of the Schr\"odinger equation
involving any (time-dependent) Hamiltonian, over any stretch of time, one can
in a simple way extend it to a closed or cyclic solution, then use
the Aharonov-Anandan method to identify a geometric phase. Both
these extensions of Berry's original framework used quantum
mechanical notions, specifically the Schr\"odinger equation. The
third step in the direction of increasing generality was taken by
 Mukunda and Simon\,\cite{nm-ap1,nm-ap2}\,: the geometric 
phase is entirely kinematical in
content, not requiring a Hamiltonian operator and associated
Schr\"odinger equation. It is determined once one is given a
(sufficiently smooth) curve of unit vectors in any complex Hilbert
space, without reference to any specifically quantum mechanical
notions. (The relevant expressions and definitions are recalled
below).

Turning to the efforts of the second kind, within quantum mechanics
we may cite the Bohm-Aharonov effect (already dealt with by Berry in
his original work), and the clarification of the connection between
Bargmann invariants and geometric phases\,\cite{bargmann}. Beyond these, it is
interesting that many instances of the geometric phase have been
identified within classical (wave) optics---the Gouy phase from
1890\,\cite{gouy1,gouy2}; 
the work by  Rytov, and 
Vladimirskii\,\cite{rytov,vladimirskii} in 1938 and
1941 on the behavior of light polarisation in the short wavelength
limit of wave optics; and Pancharatnam's studies in 1956\,\cite{pancharatnam}
 on light
polarisation and an associated phase. These classical optics examples
will be reviewed briefly in the next Section. It will be seen that
even in these situations the basic idea of a complex linear vector
space carrying a hermitian inner product, usually regarded as
characteristic of quantum mechanics, is essential to be able to
identify `classical geometric phases'. In this way, the deep link
between Bargmann invariants and geometric phases, and the
connection to Berry's original discovery, are always kept in evidence.

With this background, we now describe
briefly the kinematic approach to the geometric phase. Let
$\mathcal{H}$ be a complex Hilbert space of any dimension, with
vectors $\psi,\phi,\cdots$ and inner product $(\psi,\phi)$\,\cite{inner-product}. 
 In quantum mechanics $|(\psi,\phi)|^2$ is related to a probability; in
classical wave optics $\|\psi\|^2=(\psi,\psi)$ generally stands for
light intensity which can, but need not be, normalised. 
We next denote by $\mathcal{B}$ the unit sphere in
$\mathcal{H}$:
\begin{equation}\label{eq1.1}
\mathcal{B}=\{\psi\in\mathcal{H}\,\,|\,\,\|\psi\|^2 =
(\psi,\psi)=1\}\subset\mathcal{H}\,.
\end{equation}
The group $U(1)$ of complex phase factors acts on $\mathcal{B}$
(also on $\mathcal{H}$) in a natural way:
\begin{equation}\label{eq1.2}
\psi\in\mathcal{B}\to\psi^\prime=e^{i\alpha}\psi
\in\mathcal{B}~,~0\le \alpha < 2\pi\,.
\end{equation}
The quotient $\mathcal{B}/U(1)$, i.e., collections or equivalence classes of vectors
$\{e^{i\alpha}\psi, \psi ~\text{fixed},~0\le\alpha<2\pi\}$ differing
only by phases, forms the `ray space' $\mathcal{R}$:
\begin{equation}\label{eq1.3}
\mathcal{R}=\mathcal{B}/U(1)=\{\rho(\psi)=\psi\psi^\dag\,|\,\psi\in\mathcal{B}\}\,.
\end{equation}
Whereas $\mathcal{B}$ is a subset of $\mathcal{H}$, 
{\em the ray space ${\cal R}$ is not}\,: the
$\mathcal{B}-\mathcal{R}$ relationship is that there is a projection
map $\pi$ from the former to the latter:
\begin{equation}\label{eq1.4}
\pi:~\mathcal{B}\to\mathcal{R}~:~\psi\in\mathcal{B}\to
\pi(\psi)=\rho(\psi)\in\mathcal{R}\,.
\end{equation}
Referring to an earlier comment, $\mathcal{B}$ is a $U(1)$ principal
fibre bundle over the base $\mathcal{R}$. In the quantum mechanics
context, points in $\mathcal{R}$ correspond one-to-one to 
physical pure states.

In this framework, given any continuous piecewise once
differentiable parametrised curve $\mathcal{C}$ in $\mathcal{B}$,
\begin{equation}\label{eq1.5}
\mathcal{C} = \{\psi(s)\in\mathcal{B}\,|\, s_1\le s\le s_2\} \subset
\mathcal{B},
\end{equation}
with image $C$ in $\mathcal{R}$,
\begin{align}\label{eq1.6}
C &= \pi[\mathcal{C}] \nonumber\\
&= \{\rho(s) =
\rho(\psi(s)) = \psi(s)\psi(s)^\dag\,|\, s_1\le s \le
s_2\}\subset\mathcal{R}\,,
\end{align}
the geometric phase  $\varphi_g[C]$ is defined:
\begin{align}\label{eq1.7}
\varphi_g[C] &=\varphi_{\rm tot}[\mathcal{C}]
       -\varphi_{\rm dyn}[\mathcal{C}]\,,\nonumber\\
\varphi_{\rm tot}[\mathcal{C}] &=\arg(\psi(s_1),\psi(s_2)),\nonumber\\
\varphi_{\rm dyn}[\mathcal{C}] &=\text{Im}
\int_{s_1}^{s_2} ds \left(\psi(s), \frac{d\psi(s)}{ ds}\right)\,.
\end{align}
As indicated, $\varphi_g[C]$ is a functional of
$C\subset\mathcal{R}$, while $\varphi_{\rm tot}$ and $\varphi_{\rm dyn}$ are
both functionals of $\mathcal{C}\subset\mathcal{B}$.

An important consequence of this definition is a result involving
the so-called Bargmann invariants\,\cite{nm-ap1}. The simplest such invariant
involves three pairwise nonorthogonal vectors
$\psi_1,\psi_2,\psi_3\in\mathcal{B}$ and is the expression
\begin{equation}\label{eq1.8}
\Delta_3(\psi_1,\psi_2,\psi_3)=(\psi_1,\psi_2)(\psi_2,\psi_3)(\psi_3,\psi_1)\,,
\end{equation}
which is in general complex. The fact that $\Delta_3(e^{i\alpha_1}\psi_1,
\,e^{i\alpha_2}\psi_2,\,e^{i\alpha_3}\psi_3)$ 
 = $\Delta_3(\psi_1,\psi_2,\psi_3)$ for all real
$\alpha_1,\alpha_2,\alpha_3$ shows that $\Delta_3$ 
lives in ${\cal R}$ rather than in ${\cal B}$.
Indeed Bargmann, during the course of his famous proof of
Wigner's theorem, introduced $\Delta_3$ simply to point out this
gauge-invariance property and to indicate that it could
be used to distinguish between unitaries and antiunitaries\,:
while $\Delta_3$ is invariant under unitaries, its argument changes
signature under antiunitaries. It is in \cite{nm-ap1,nm-ap2} that this object
introduced by Bargmann almost in passing was elevated
to become the basis of a complete kinematic theory
of geometric phase.

To relate arg$\left(\Delta_3(\psi_1,\psi_2,\psi_3)\right)$ to a 
geometric phase, it is
necessary to connect the `vertices' $\psi_1,\psi_2,\psi_3$ pairwise
in some way, so as to construct a closed continuous piecewise
once-differentiable loop in $\mathcal{B}$  reminiscent of cyclic
quantum evolution. This can be done using the idea of geodesics in
$\mathcal{R}$. Given two nonorthogonal vectors
$\psi,\phi\in\mathcal{B}$, and assuming for definiteness that
$(\psi,\phi)$ is real positive, the (shorter) geodesic in
$\mathcal{R}$ connecting $\rho(\psi)$ to $\rho(\phi)$ is the image
under $\pi$ of the curve $\mathcal{C}=\{\psi(s)\}\subset\mathcal{B}$
described as follows:
\begin{align}\label{eq1.9}
(\psi,\phi)&=\cos\theta\,,\quad0<\theta<\pi/2\,:\nonumber\\
\psi(s)&=\psi\cos\,s +\phi^\perp \frac{\sin\,s}{\sin\theta}\,,\quad
0\le s\le\theta,\,\nonumber\\
\phi^\perp &=\phi-\psi\cos\theta\,.
\end{align}
Along this $\mathcal{C}$ one has
\begin{align}\label{eq1.10}
(\psi(s^\prime),\psi(s))=\cos(s^\prime-s)\,,\quad0\le s^\prime,
s\le\theta\,.
\end{align}
Then the connection between Bargmann invariants and geometric
phases is:
\begin{align}\label{eq1.11}
&\arg\left(\Delta_3(\psi_1,\psi_2,\psi_3)\right)=
-\varphi_g[C],\,\nonumber\\
C&=\text{triangle in $\mathcal{R}$ with vertices
$\rho(\psi_1),\rho(\psi_2),\rho(\psi_3)$}\nonumber\\
&~~~~~~~\text{and connecting geodesics as sides.}
\end{align}
A very far-reaching generalisation of this relation, when $\dim
\mathcal{H}\ge 3$, has been developed more recently\,\cite{bi-1,bi-2,bi-3,bi-4}.

 It is worth emphasizing that the framework described above, based
on the triplet of spaces $\mathcal{H},\mathcal{B},\mathcal{R}$,
supports the geometric phase concept in a simple and direct way.
Though suggested by the formal ({complex linear space}) structure 
of quantum mechanics, it
can be used in other situations as well, such as classical wave
optics. We adopt this viewpoint in trying to define geometric phases
in various physical,  particularly classical optical, situations.

Before we outline the organization of the 
material of this paper, it may be useful to add an extended 
 remark by way of pointing to the precise context of this work.
 There has been considerable interest in recent times to  
understand the interplay between the spatial degree of freedom (coherence)  
and polarisation degree of  {\em electromagnetic} 
beams [see, for instance, \cite{massimo} and references therein]. 
 It is equally important to understand the behaviour of this 
interplay as the Maxwell beam passes through an optical system.
Indeed, it turns out that the {\em defining} properties of 
the age old Mueller matrix cannot be correctly enumerated without 
consideration of this interplay or entanglement\,\cite{mueller1,mueller2}.
  
 A lens of focal length $f$ relates the 
 output field amplitude $\psi_{\rm out}(x_1,\,x_2)$ 
(just after the  lens plane) to the 
input $\psi_{\rm in}(x_1,\,x_2)$ (just before the lens plane), 
$(x_1,\,x_2)$ being Cartesian variables in the transverse plane, 
through  
\begin{align}
\psi_{\rm out}(x_1,\,x_2) = 
 \exp \left(-i \frac{x_1^{\,2} + x_2^{\,2}}{2\lbar f} \right)
\psi_{\rm in}(x_1,\,x_2).   
\end{align}
But when it comes to vector waves, it is clear that the same 
transformation applied to  every 
(Cartesian) component of the electric field vector $\boldsymbol{E}(x_1,\,x_2)$ 
 will not map solutions
of Maxwell equations at the input plane to solutions 
at the output, for such a democratic action 
 on the electric field components {\em does not} respect the 
transversality  condition $\boldsymbol{\nabla \cdot} \boldsymbol{E} = 0$. 
 Since this condition is a {\em constraint} connecting the spatial degrees of 
freedom to the polarization degree, it would be respected 
only if   the spatial modulation 
$\exp \left(-i \frac{x_1^{\,2} + x_2^{\,2}}{2\lbar f}\right)$ 
is accompanied by `appropriate' {\em local rotations} of the 
electric field components (local polarization)\,\cite{front}. 

Let us arrange the components of the electric and magnetic field 
amplitude vectors $\boldsymbol{E}(x_1,\,x_2)$, $\boldsymbol{B}(x_1,\,x_2)$ 
in a transverse plane $z$ = constant into a six-component electromagnetic vector 
\begin{align}
\boldsymbol{\Lambda}(x_1,\,x_2) 
=\left(\begin{array}{c}\boldsymbol{E}(x_1,\,x_2)\\ 
\boldsymbol{B}(x_1,\,x_2) \end{array}\right).
\end{align}
The approach of\,\cite{front} rooted at the very Poincar\'{e} symmetry 
of the Maxwell system of equations led to this fundamental result: 
 if $T(x_1,\,x_2)$ is the amplitude transmittance function of an optical system 
in scalar Fourier optics [\,a lens, for instance, has 
$T(x_1,\,x_2) = \exp \left(-i \frac{x_1^{\,2} + x_2^{\,2}}{2\lbar f}\right)$\,], 
then  the action 
\begin{align}
T:~ &\boldsymbol{\Lambda}_{\rm in}(x_1,\,x_2) \to 
\boldsymbol{\Lambda}_{\rm out}(x_1,\,x_2) 
 = T(Q_1,\,Q_2)\boldsymbol{\Lambda}_{\rm in}(x_1,\,x_2),\nonumber\\
&~~Q_1 = x_1\,1\!\!1_{6\times6}\,+\,\lbar G_1,~~~
Q_2 = x_2\,1\!\!1_{6\times 6}\,+\,\lbar G_2,
\end{align}
where $G_1,\,G_2$  are a pair of $6\times6$ {\em numerical} matrices arising from 
the structure of the Poincar\'{e} group\,\cite{front}, 
{\em does take} solutions $\boldsymbol{\Lambda}_{\rm in}(x_1,\,x_2)$ of 
Maxwell's equations to solutions $\boldsymbol{\Lambda}_{\rm out}(x_1,\,x_2)$. 
That is, the matrices $G_1,\,G_2$ effect 
on the components of $\boldsymbol{\Lambda}(x_1,\,x_2)$ 
the correct local rotations alluded to above\,\cite{front}. 
 This result readily leads to Fourier optics for 
Maxwell beams\,\cite{fourier} and to electromagnetic Gaussian 
beams\,\cite{gaussian}, resulting in a straight forward description of 
not only the longitudinal component but also the cross-polarisation 
component\,\cite{cross}. It is in respect of this result that 
the late Henri Bacry 
anticipated: ``it is highly probable that a rigorously gauge theory 
will be developed in a near future'', the local rotations referred to 
above constituting ``an $SO(3)$ gauge group''\,\cite{bacry}.  

The work presented here is only the first step of an ambitious programme 
 constituting our attempt towards a possible realization of this anticipation.
 While the earlier formulation of Fourier optics for Maxwell 
beams\,\cite{fourier} concentrated on paraxial propagation about 
a {\em fixed direction}, the present  work aims at laying a 
global and structurally robust 
skeleton in the {\em space of directions}, 
 handling satisfactorily the  well known topological 
obstructions. In the sequel, we plan to adapt suitably 
 the methods of \cite{fourier} in local patches of  the 
space of directions, and then `stitch' together the patches  
 in a smooth manner to arrive at the general case.  

The contents of this paper are organised as follows. 
Section~II gives brief accounts of three applications of the 
geometric phase concept to classical optical situations\,: the Gouy phase 
in scalar paraxial wave optics; the Pancharatnam study of 
phases in pure polarization  optics with fixed propagation direction; 
and the behaviour of polarisation in the eikonal or ray limit of 
 Maxwell's equations in a transparent medium with given refractive index function. 
 The Pancharatnam case uses the Poincar\'e sphere $S^2_{\rm pol}$
 of polarisation states, for a fixed direction of propagation, 
while the ray case uses the sphere of propagation directions 
$S^2_{\rm dir}$. In all these cases, the use of the basic quantum mechanical 
framework is highlighted.  Section~III builds on the last example of Section~II 
in two ways---the generalisation from the unphysical case of a single ray 
to a physical beam of finite cross-sectional area made up of 
a narrow bundle of nearly  parallel rays; and the inclusion of polarization gadgets 
in the path of the beam. Once again the quantum mechanical framework 
proves adequate, and now both spheres 
$S^2_{\rm pol}$, $S^2_{\rm dir}$ come into the picture. 
Section~IV takes up certain global features of the sphere of directions 
$S^2_{\rm dir}$, and builds on a recent suggestion\,\cite{rnss} that 
passage to the  complex extension of the tangent planes to 
$S^2_{\rm dir}$ removes an obstruction which exists in the real domain.
(The work of\,\cite{rnss} was motivated in part by earlier 
works of\,\cite{bhandari89,hannay98,tavrov00}.) 
 Using elementary group theoretical arguments, based on the groups $SO(3)$ 
and $SU(2)$, a particularly simple global basis of complex orthonormal 
vector fields  tangent to  $S^2_{\rm dir}$ is constructed.  
Section~V uses the constructions of Section~IV to study 
again the beams of Section~III and their geometric phases\,:
 in appropriate situations, the complete geometric phase separates into 
a contribution from   $S^2_{\rm pol}$  and another from 
$S^2_{\rm dir}$. The final Section~VI contains some concluding remarks, 
while the Appendix compares the present framework for handling geometric phases 
with that proposed in \cite{rnss}.

\section{Examples of classical optical geometric phases}
In this Section we review three situations in classical optics
displaying geometric phases, presenting only the essential details.
The first concerns scalar wave optics, the other two include
polarisation. Quantum mechanical notation is used when convenient\,\cite{simonjosa00}.

\subsection{The case of the Gouy phase}
We deal with the scalar optical wave field in free space, with fixed
(angular) frequency $\omega$, wave number $k=\omega/c$ and wavelength
$\lambda=2\pi/k$. In the paraxial approximation to the Helmholtz
equation, with the positive $z$-axis as the propagation direction,
we obtain the paraxial wave equation in two transverse dimensions\,: 
\begin{equation}\label{eq2.1}
i\lbar\frac{\partial}{\partial z}\psi(x,y;z) =-\frac{\lbar^2}{2}\left(\frac{\partial^2}{\partial x^2}+
\frac{\partial^2}{\partial y^2}\right)\psi(x,y;z)\,,
\end{equation}
where $\lbar={\lambda/(2\pi)}=k^{-1}$, reminiscent of $\hbar$\,.
(The exponential factor $e^{i(kz-\omega t)}$ has been omitted in
$\psi$). This is formally similar to the Schr\"odinger equation in
quantum mechanics for a free nonrelativistic particle of unit mass in
two dimensions, with $\lbar$ in place of $\hbar$, and with the longitudinal
variable z playing the role of `time'. The
``Hamiltonian operator'' $H$ for Eq.\,(\ref{eq2.1}) is
\begin{equation}\label{eq2.2}
H=\frac{1}{2}\left(p_x^2+p_y^2\right),\;\;
p_x=-i\lbar\frac{\partial}{\partial x}\,,\;
p_y=-i\lbar\frac{\partial}{\partial y}\,.
\end{equation}
If we restrict to one transverse dimension, we have the simpler
paraxial wave equation
\begin{equation}\label{eq2.3}
i\lbar\frac{\partial}{\partial z}\psi(x;z)=H\psi(x;z)\,,\;
H=\frac{1}{2} p_x^2=-\frac{\lbar^2}{2}\frac{\partial^2}{\partial x^2}\,.
\end{equation}

Our focus is on `centred' Gaussian solutions to this paraxial wave equation, and
their phases. Based on group theoretical considerations, it is
convenient to parametrise normalised centred Gaussians by a complex
variable $q$ with negative imaginary part, i.e., lying in the
lower half complex plane:
\begin{align}\label{eq2.4}
\text{Im}~q<0~:~&\psi_0(x;q)=
\left(\frac{-\text{Im}~q}{\pi\lbar|q|^2}\right)^{1/4}
\exp \left(\,i \,\frac{x^2}{2\lbar q}\,\right),\nonumber\\
&\int_{-\infty}^\infty dx~|\psi_0(x;q)|^2=1\,.
\end{align}
Then the centred Gaussian solution to Eq.\,(\ref{eq2.3}), with width
$w$ in the `waist' plane  $z=0$, is:
\begin{align}\label{eq2.5}
\psi(x;0)&=\psi_0 \left(x;-i\frac{w^2}{2\lbar}\right) = 
\left(\frac{2}{\pi w^2}\right)^{1/4}\exp \left( - \frac{x^2}{ w^2}\right)\nonumber\\
&\longrightarrow \psi(x;z)
      = e^{i\varphi_G (z)}\psi_0\left(x; q(z)\right);\nonumber\\
\varphi_G (z)&= - \frac{1}{ 2}\,\tan ^{-1}\left(\frac{z}{ z_R}\right),~~ 
q(z)= z-iz_R,\nonumber\\
z_R &= \text{Rayleigh range}\; = w^2/2\lbar = \pi w^2/\lambda. 
\end{align}
Here, $\varphi_G(z)$ is the evolving Gouy phase. It is the
argument of $\psi(0;z)$ (on-axis phase at the plane $z =$ constant) 
and `jumps' by $-\pi/2$ (by $-n\pi/2$ for $n$ transverse dimensions)
across the waist plane\,: 
\begin{align}\label{eq2.6}
&\varphi_G(z)=\arg \psi(0;z),\nonumber\\
&\varphi_G(\infty)-\varphi_G(-\infty) =-\pi/2\,.
\end{align}
That the parameter $q$ lives in the lower half plane is a consequence
of our taking monochromatic time-dependence
in the usual form $\exp ( -i \omega t)$. Had it been taken in the
form $\exp(i\omega t)$, as some authors do, then $q$ would live in
the upper half plane. The evolution of Gaussian beams
through first order systems described by abcd-matrix is
governed by the well known Kogelnik abcd-law\,\cite{kogelnik65},
\begin{align}\label{eq2.6b}
q_{\rm in} \to q_{\rm out} =
\frac{a\, q_{\rm in} + b}{c\, q_{\rm in} + d},
\end{align}
of which the particular case $q(z_1) \to q(z_2) = q(z_1) +
(z_2-z_1)$, corresponding to free propagation from $z_1$ to
$z_2$  [\,i.e., $(a, b, c, d) = (1, z_2-z_1, 0, 1)$\,], is already quoted
in Eq.\,\eqref{eq2.5}. It may be noted in passing that the abcd-law
has been generalized to partially coherent Gaussian
beams, the so-called Gaussian Schell-model beams, 
in\,\cite{simon84} and to arbitrary beams in\,\cite{simon88}.

Our aim now is to show that $\varphi_G(z)$ is essentially a
geometric phase. For this we need the extension of the relation
(\ref{eq1.11}) to the four-vertex Bargmann invariant, and then
specialize it in a particular way. For the moment we use quantum
mechanical notation, with $\psi$ denoting a Hilbert space vector.
The generalisation of the connection (\ref{eq1.11}) is
\,[$\psi_1,\cdots,\psi_4$ are unit vectors\,]:
\begin{align}\label{eq2.7}
&\Delta_4(\psi_1,\psi_2,\psi_3,\psi_4)
=(\psi_1,\psi_2)(\psi_2,\psi_3)(\psi_3,\psi_4)(\psi_4,\psi_1)\,,\nonumber\\
&\;\;\arg~\Delta_4(\psi_1,\psi_2,\psi_3,\psi_4)=-\varphi_g[C]\,,\nonumber\\
C&=\text{quadrilateral in $\mathcal{R}$ with vertices
$\rho(\psi_1),\cdots,\rho(\psi_4)$}\nonumber\\
&~~~~~~\text{and geodesics connecting}\nonumber\\
&~~~~~~~~  \text{$\rho(\psi_1)$ to
$\rho(\psi_2),\,\cdots,\,\rho(\psi_4)$ to $\rho(\psi_1)$ as sides.}
\end{align}
Now to the specialisation of this relation. Let $s$ be an evolution
parameter, and $H_0$ a `Hamiltonian operator' independent of $s$; 
and let $\psi_0(s)$ obey the `Schr\"odinger equation'
\begin{align}\label{eq2.8}
i\frac{d}{ ds} \psi_0(s)=H_0\psi_0(s),
\end{align}
so that
\begin{align}\label{eq2.9}
\psi_0(s_2)=e^{-i(s_2-s_1)H_0}\psi_0(s_1)\,.
\end{align}
In the relation (\ref{eq2.7}) we now choose $\psi_1$ to be a
convenient `reference vector' $\psi_R$, which allows the measurement
of the phase $\varphi(s)$ of $\psi_0(s)$ with respect to it in the Pancharatnam
sense (i.e., through an inner-product)\,:
\begin{align}\label{eq2.10}
\varphi(s)=\arg(\psi_R, \psi_0(s))\,.
\end{align}
Further, we choose $\psi_2=\psi_0(s_1), \psi_3=$ a `zero energy' vector
$\psi_E$ obeying $H_0\psi_E=0$, and $\psi_4=\psi_0(s_2)$. Then,
using also Eq.\,(\ref{eq2.9}), the connection (\ref{eq2.7}) becomes:
\begin{align}\label{eq2.11}
\varphi(s_2)-\varphi(s_1)&=\varphi_g\text{[\,quadrilateral in
$\mathcal{R}$ with vertices}\nonumber\\
&~~~~~\rho(\psi_R),\rho(\psi_0(s_1)), \rho(\psi_E),\rho(\psi_0(s_2)),\nonumber\\
&~~~~~~~\left.\text{and geodesic sides}\,\right].
\end{align}
To apply Eq.\,(\ref{eq2.11}) to the present case, we set
$s=\lbar^{-1}z$ and $H_0=\frac{1}{2}p_x^2$, so (\ref{eq2.8}) becomes
(\ref{eq2.3}); and associate `wave functions' $\psi_R(x), \psi(x;z),
\psi_E(x)$ with the Hilbert space vectors $\psi_R, \psi_0(s),
\psi_E$ respectively. To arrange $\varphi(s)$ in Eq.\,(\ref{eq2.10})
to be the Gouy phase $\varphi_G(z)$, recalling the `on-axis'
identification in Eq.\,(\ref{eq2.6}), the wave function $\psi_R(x)$
must become essentially $\delta(x)$. Next, to obey the condition
$H_0\psi_E=0$ the wave function $\psi_E(x)$ must become
$x$-independent, 
i.e., a plane wave with wave vector strictly along the z-axis 
(recall that we have dropped, following \eqref{eq2.1}, 
a factor $\exp[i(kz-\omega t)]$). With these clues we take $\psi_R(x)$ and
$\psi_E(x)$ to be particular limiting forms of $\psi_0(x; q)$ (and
as our interest is in phases alone we disregard real factors which
diverge or vanish in the limits):
\begin{align}\label{eq2.12}
\psi_R(z)&=\lim_{q_1=0,q_2\to0_-} \psi_0(x;
q_1+iq_2)\nonumber\\
 &=\lim_{q_2\to0_-} \frac{1}{(-\pi\lbar q_2)^{1/4}}
\exp\left(\frac{x^2}{ 2\lbar q_2}\right) \sim \delta(x)\,;\nonumber\\
\psi_E(x)&=\lim_{q_1\to 0,q_2\to -\infty} \psi_0(x;
q_1+iq_2)\nonumber\\
 &=\lim_{q_2\to -\infty} \frac{1}{(-\pi\lbar q_2)^{1/4}}
\exp\left(\frac{x^2}{2\lbar q_2}\right) \sim \text{constant in}~x\,.
\end{align}
Then indeed, with $s=z/\lbar$,
\begin{align}\label{eq2.13}
\varphi(s)=\arg(\psi_R,\psi_0(s))&=\arg\left(\int_{-\infty}^\infty
dx ~\psi_R(x)^*\psi(x; z)\right)\nonumber\\
&=\arg\psi(0; z)=\varphi_G(z)\,,
\end{align}
so Eq.\,(\ref{eq2.11}) becomes:
\begin{align}\label{eq2.14}
&\varphi_G(z_2)-\varphi_G(z_1)\nonumber\\
&~~~=\varphi_g[\,\text{quadrilateral in
$\mathcal{R}$, vertices $\rho(\psi_R(x) \sim \delta(x))$},\nonumber\\
&~~~~~~~~~~\rho(\psi(x;z_1)),\rho(\psi_E(x) \sim \text{constant})\,, 
 \rho(\psi(x;z_2))\nonumber\\
&~~~~~~~~~~~~~\text{and geodesic sides}\,]\,.
\end{align}
This already shows that (differences of) Gouy phases are certain geometric
phases. However, for improved understanding, we can analyse the
right hand side further as follows.

The argument of $\varphi_g$ on the right hand side in the result
(\ref{eq2.14}) is a quadrilateral in the `ray space' $\mathcal{R}$,
with geodesic sides, as it should be. The geodesics needed here are
to be constructed in the manner of Eq.\,(\ref{eq1.9}) at the vector
space or wave amplitude level, followed by projection $\pi$ to
$\mathcal{R}$. A quite subtle analysis\,\cite{comment-rabei}
 (here omitted) shows that in
the present instance (and some others of interest) we can use
`geodesics' drawn within the manifold of centred Gaussian
amplitudes (which may be called `constrained geodesics'), and the
basic connection (\ref{eq1.11}), (\ref{eq2.7}) between Bargmann
invariants and geometric phases continues to be valid. Next, as the
definition (\ref{eq1.7}) of geometric phases shows, for practical
calculations one can choose any convenient `lift' $\mathcal{C}$ of
$C$ at the level of Hilbert space vectors or wave amplitudes,
obeying $C=\pi[\mathcal{C}]$. In particular if $\mathcal{C}$ is chosen
to be a closed loop (the quadrilateral ${C}$ in $\mathcal{R}$
is of course closed) the piece $\varphi_{\rm tot}[\mathcal{C}]$ in
Eq.\,(\ref{eq1.7}) vanishes and we are left with
$\varphi_g[C]=-\varphi_{\rm dyn}[\mathcal{C}]$. Beyond this, one can
use the phase freedom at each point along $\mathcal{C}$ to assume,
in the present case, that $\mathcal{C}$ is a closed loop within the
space of centred Gaussian wave functions $\psi_0(x; q)$. It can
then be pictured or drawn as a closed curve in the lower half of the
complex $q$ plane. One must only ensure that the `vertices' are
chosen properly, so as to project onto the vertices specified in
$\mathcal{R}$ in (\ref{eq2.14}), and the connecting curves represent
`constrained geodesics' properly. When all this is done, the result
is as shown in Fig.\,1.
\begin{figure}
\begin{center}
\scalebox{0.6}{\includegraphics{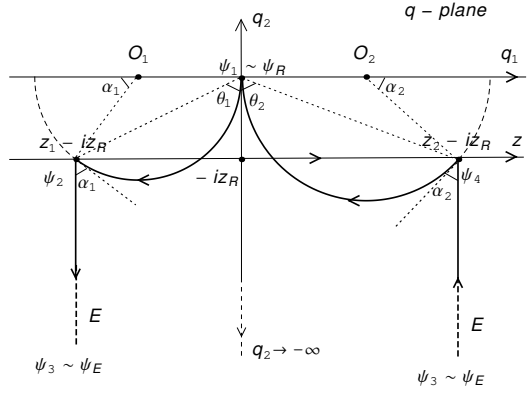}}
\end{center}
\caption{Illustrating the hyperbolic geometry of the lower half complex 
$q$-plane 
underlying Gaussian beams and the abcd-law. Free propagation corresponds to the 
horizontal line passing through $q=-iz_R$. 
The two circular geodesics 
are centred at $O_1,\,O_2$. For the geodesic quandrilateral 
$\psi_1 \to \psi_2 \to \psi_3 \to \psi_4 \to \psi_1$ the angles at 
$\psi_1,\, \psi_3$ vanish, while the angles at $\psi_2,\,\psi_4$ 
become $\pi/2$ at $z_1= -z_R$ and $z_2= z_R$ respectively. Thus 50\% of the total 
Gouy phase `jump' occurs within a propagation distance  $2z_R$ around the waist, 
$z_R$ decreasing quadratically with decreasing waist size $w$.}
\end{figure}

The arcs connecting $q=0$ to
$q=-iz_R+ z_1$ (i.e., $\psi_1$ to $\psi_2$) and
$q=-iz_R+ z_2$ to $q=0$ (i.e., $\psi_4$ to $\psi_1$) are both
circular, with centres on the $q_1$ axis. The straight lines 
connecting $q=-iz_R+ z_1$ to $q 
= -i\infty$ (i.e., $\psi_2$ to $\psi_3$) and $q=-i\infty$ to $q 
= -iz_R +  z_2 $    (i.e., $\psi_3$ to $\psi_4$) are
both vertical, parallel to the $q_2$ axis. All of them
taken in sequence `represent' the closed $\mathcal{C}$:
\begin{align}\label{eq2.15}
\mathcal{C} \sim q=0 &\rightarrow q=-iz_R+ z_1 \rightarrow
q=-i\infty \nonumber\\
 &~~~\rightarrow q = - iz_R+ z_2 \to q=0
\end{align}
and Eq.\,(\ref{eq2.14}) takes the more explicit form
\begin{align}\label{eq2.16}
\varphi_G(z_2)&-\varphi_G(z_1)= -\varphi_{\rm dyn}[\mathcal{C}]\nonumber\\
&~~ = -\text{Im} \oint_{\mathcal{C}} \left\{ \text{\Large{(}}\psi_0(x;q),
\frac{\partial}{\partial q_1} \psi_0(x;q) \text{\Large{)}}\, dq_1\right.\nonumber\\
&~~~~~~~~~~\left.+\text{\Large{(}}\psi_0(x;q), 
\frac{\partial}{\partial q_2}\psi_0(x;q)\text{\Large{)}}\,
dq_2\right\}.
\end{align}
The integration in the $q$ half plane is along the curve
(\ref{eq2.15}), while the inner products in Hilbert space
$\mathcal{H}=L^2(\mathbb{R})$ (integrations with respect to $x$) are
left implicit. With some effort one can confirm that the integral over
$\mathcal{C}$ on the right in Eq.\,(\ref{eq2.16}) indeed reproduces
the difference between Gouy phases on the left, as determined by
Eq.\,(\ref{eq2.5}).

Alternatively, the line integral in Eq.\,\eqref{eq2.16} equals\,\cite{gouy-rsnm}
the negative of one-fourth of the (hyberbolic) area of
the enclosed quadrilateral, the abcd-law being a signature
of the natural Lobachevskian hyperbolic geometry
with metric
\begin{align}\label{eq2.16b}
d\ell^2 = (dq^2_1 + dq^2_2)/q_2^2\,, 
\end{align}
underlying the manifold of Gaussian states, the lower
half $q$-plane. The area itself is given by the `hyperbolic
deficiency' which, for a (geodesic) quadrilateral, equals
$2\pi$ minus sum of the interior angles.

The interior angle vanishes at $R$ as well as at $E$. For
the other two angles $\alpha_1, \alpha_2$ we see from Fig.\,1 that $\alpha_j =
2(\pi/2-\theta_j)$ and $\tan{\theta_j} = z_j/z_R$. Thus the deficiency
equals $2(\theta_1+\theta_2)$, leading to a geometric phase of $-(\theta_1+
\theta_2)/2$. One thus recovers
\begin{align}\label{eq2.16c}
\varphi_G(z_2)-\varphi_G(z_1) = -\frac{1}{2}
({\rm arctan}(z_2/z_R) - {\rm arctan}(z_1/z_R))\,,
\end{align}
well known in the context of laser beams, now as a geometric
phase.

We appreciate that this demonstration of the link between Gouy
and geometric phases is fully within the
$\mathcal{H}-\mathcal{B}-\mathcal{R}$ framework of quantum mechanics
used in the quantum kinematic approach\,\cite{nm-ap1,nm-ap2}
to  geometric phase, briefly recapitulated in Section~1.

\subsection{The Pancharatnam case}
Now we include the polarization degree of freedom, and to begin with
consider the extreme case when it is the only variable. With given
frequency $\omega$ and wave number $k=\omega/c$, we {\em fix also the
direction of propagation} to be the positive $z$-axis, and consider
plane waves in various states of pure polarization. The analysis
again falls perfectly into the quantum mechanical 
$\mathcal{H}-\mathcal{B}-\mathcal{R}$ scheme. 
Dropping the standard factor $e^{i(kz-\omega t)}$, 
at each $z$ the electric field is a complex two-component
vector in the transverse x-y plane,
$\boldsymbol{E}=\binom{E_1}{E_2}$. The Hilbert space $\mathcal{H}$
for this case is then $\mathcal{H}=\mathbb{C}^2$ of dimension two.

As is well known, the spaces $\mathcal{B}$ and $\mathcal{R}$ are
$S^3$ and $S^2_{\rm pol}$ respectively, unit spheres in real four 
and three-dimensional Euclidean spaces, the latter being the Poincar\'e
sphere of pure polarization states\,:
\begin{align}\label{eq2.17}
\mathcal{H}=\mathbb{C}^2 \to \mathcal{B}_3 = S^3 \xrightarrow{\boldsymbol{\pi}}
\mathcal{R}_2 = \mathcal{B}_3/U(1) = S^2_{\rm pol}\,.
\end{align}
Given $\boldsymbol{E}$ at some $z$, the corresponding pure
polarization state is represented by a point $\hat{\boldsymbol{n}}\in S^2_{\rm pol}$
computed as follows:
\begin{align}\label{eq2.18}
\boldsymbol{E} \to ~\hat{\boldsymbol{n}} &= (\boldsymbol{E}^\dag
\boldsymbol{E})^{-1} ~\boldsymbol{E}^\dag \boldsymbol{\tau} \boldsymbol{E}
\in S^2_{\rm pol},\,\nonumber\\
& ~~~~ ~~~\boldsymbol{\tau} = (\tau_1,\tau_2,\tau_3) 
  = (\sigma_3, \sigma_1, \sigma_2),
\end{align}
where the $\sigma$'s are the standard quantum mechanical Pauli
matrices. Under free propagation governed by the free Maxwell
equations, the amplitude $\boldsymbol{E}$ and the polarization state
$\hat{\boldsymbol{n}}$ are both constant: $\hat{\boldsymbol{n}}$ 
is stationary on $S^2_{\rm pol}$.

More generally, we imagine the plane wave passing through
transparent linear intensity preserving polarization gadgets which
act on $\boldsymbol{E}$ and alter the polarization state $\hat{\boldsymbol{n}}$.
These are placed at various locations (lumped) or over various 
stretches (distributed) along
the $z$-axis, separated from one another by intervals of free
propagation. The effect of such gadgets on $\boldsymbol{E}$ is again
governed by Maxwell's equations for propagation of the field
through suitable transparent material media. As our interest is only
in the behaviour of the polarization state $\hat{\boldsymbol{n}}$, the intensity
being held constant, we can represent each polarization gadget by a
corresponding element of the two-dimensional unitary unimodular
group $SU(2)$\,\cite{comment-su(2)}\,---the additional $U(1)$ phase 
in the full unitary
group $U(2)$ is not relevant for this purpose. With this physical
picture in place, let us write $\boldsymbol{E}(z)$ and $\hat{\boldsymbol{n}}(z)$
for the field and the polarization state at position $z$ along the
propagation axis:
\begin{align}\label{eq2.19}
\boldsymbol{E}(z)\in\mathbb{C}^2\to \hat{\boldsymbol{n}}(z)
=(\boldsymbol{E}(z)^\dag\boldsymbol{E}(z))^{-1} \,\boldsymbol{E}(z)^\dag
\boldsymbol{\tau}\boldsymbol{E}(z) 
\in S^2_{\rm pol}\,.
\end{align}
Then $\boldsymbol{E}(z)$ evolves according to the Schr\"odinger-like
equation
\begin{align}\label{eq2.20}
i \,\frac{d\boldsymbol{E}(z)}{ dz} = H(z)\boldsymbol{E}(z)~,~~~H(z) =
\frac{1}{2}\boldsymbol{\tau}\cdot\boldsymbol{a}(z)~,
\end{align}
where $\boldsymbol{a}(z)$ is a real three-dimensional vector and
$H(z)$ is the `Hamiltonian'. Correspondingly for $\hat{\boldsymbol{n}}(z)$ we
have
\begin{align}\label{eq2.21}
\frac{d\hat{\boldsymbol{n}}(z)}{ dz} 
= \boldsymbol{a}(z)_\wedge \hat{\boldsymbol{n}}(z).
\end{align}
Thus while $\boldsymbol{E}(z)$ undergoes a gradually unfolding
$SU(2)$ transformation,  $\hat{\boldsymbol{n}}(z)$ experiences a gradual rotation
belonging to $SO(3)$\,\cite{comment-so(3)}. (Free propagation stretches correspond to
$H(z)=0$, and hence to $\boldsymbol{a}(z)=0$; 
for lumped elements like a quarter or half wave plate 
$\boldsymbol{a}(z)$ is a Dirac delta function). Over a finite stretch $z_1$ to $z_2$,
we have:
\begin{align}\label{eq2.22}
\boldsymbol{E}(z_2) &= \mathcal{U}(z_2,z_1)\boldsymbol{E}(z_1)~,\quad
 \mathcal{U}(z_2,z_1) \in SU(2)\,;\nonumber\\
\hat{\boldsymbol{n}}(z_2)  &= \mathcal{R}(z_2,z_1)\hat{\boldsymbol{n}}(z_1)~,\quad
\mathcal{R}(z_2,z_1) \in SO(3)\,,
\end{align}
with $\mathcal{U}(z_2,z_1)$ determining $\mathcal{R}(z_2,z_1)$ through
the well known $SU(2)\to SO(3)$ homomorphism\,\cite{comment-homo}.

If $H(z)$ and $\boldsymbol{a}(z)$ are constant from $z_1$ to $z_2$,
say $H$ and $\boldsymbol{a}$ respectively, we have
\begin{align}\label{eq2.23}
\mathcal{U}(z_2,z_1) &= e^{-i(z_2-z_1)H}~,\nonumber\\
\mathcal{R}(z_2,z_1) &=
\mathcal{R}(\hat{\boldsymbol{a}},~(z_2-z_1)|\boldsymbol{a}|)\,,\quad
\hat{\boldsymbol{a}}=\boldsymbol{a}/|\boldsymbol{a}|~,
\end{align}
where $\mathcal{R}(\hat{\boldsymbol{a}},\alpha)$ is the right handed rotation
about axis $\hat{\boldsymbol{a}}$ by amount $\alpha$\,\cite{comment-so(3)}. 
Then over such a stretch
\begin{align}\label{eq2.24}
\boldsymbol{E}(z)^\dag H \boldsymbol{E}(z) =
\frac{1}{2}\boldsymbol{E}(z)^\dag \boldsymbol{E}(z)\,\boldsymbol{a}\cdot
\hat{\boldsymbol{n}}(z) = \text{constant}\,,
\end{align}
and as $\boldsymbol{E}(z)^\dag\boldsymbol{E}(z)=$ constant as well, we
see that $\hat{\boldsymbol{n}}(z)$ moves on a latitude circle in a plane
perpendicular to $\boldsymbol{a}$. In case
$\boldsymbol{a}\cdot\hat{\boldsymbol{n}}(z)=0$, 
$\hat{\boldsymbol{n}}(z)$ moves on the great
circle perpendicular to $\hat{\boldsymbol{a}}$, 
the equator with respect to $\hat{\boldsymbol{a}}$; and as then
$\boldsymbol{E}(z)^\dag H\boldsymbol{E}(z)=0$, \textit{such stretches
contribute zero dynamical phases}.

A cyclic evolution in this Pancharatnam situation carries the
electric field over some curve $\mathcal{C}\subset \mathcal{B}_3 =S^3$
(assuming for simplicity $\boldsymbol{E}(z)^\dag\boldsymbol{E}(z)=1$),
say from $\boldsymbol{E}^{(1)}$ at $z_1$ to
$\boldsymbol{E}^{(2)}=e^{i\theta}\boldsymbol{E}^{(1)}$ at $z_2$. Then
$\hat{\boldsymbol{n}}(z)$ describes a closed loop $C_{\rm pol} 
\subset S^2_{\rm pol}$. By
Eq.\,(\ref{eq1.7}), the associated geometric phase can be readily
computed, and it turns out to be very simply related to the geometry
of $S^2_{\rm pol}$:
\begin{align}\label{eq2.25}
\varphi_g[C_{\rm pol}] &=
\varphi_{\rm tot}[\mathcal{C}]-\varphi_{\rm dyn}[\mathcal{C}]\nonumber\\
&= \arg \boldsymbol{E}^{(1)^\dag} \boldsymbol{E}^{(2)}-\text{Im}
\int_{z_1}^{z_2} dz\,\boldsymbol{E}(z)^\dag \frac{d\boldsymbol{E}(z)}{
dz}\nonumber\\
&= \theta +   \int_{z_1}^{z_2} dz\,
\boldsymbol{E}(z)^\dag H(z)\boldsymbol{E}(z)\nonumber\\
&= \frac{1}{2} \Omega[C_{\rm pol}]\,,
\end{align}
where $\Omega[C_{\rm pol}]$ is the solid angle (with sign) subtended by
$C_{\rm pol}$ at the origin of $S^2_{\rm pol}$.

In the original Pancharatnam
analysis, $C_{\rm pol}$ is a spherical triangle on $S^2_{\rm pol}$ with sides being
great circle arcs (\,i.e., geodesics\,)\,\cite{comment-arcs}, 
leading as mentioned above to
$\varphi_{\rm dyn}[\mathcal{C}]=0$ if piecewise constant  `Hamiltonians'
are used. And $\varphi_g[C_{\rm pol}]$ reduces to the negative of the
phase of a three-vertex Bargmann invariant, a special simple
instance of Eq.\,(\ref{eq1.11}): if fields $\boldsymbol{E}^{(1)},
\boldsymbol{E}^{(2)}, \boldsymbol{E}^{(3)}$ lead via Eq.\,(\ref{eq2.18})
to the vertices $\hat{\boldsymbol{n}}_1,\hat{\boldsymbol{n}}_2,
\hat{\boldsymbol{n}}_3$ of  $C_{\rm pol}$ then 
\begin{align}\label{eq2.26}
&C_{\rm pol}  =\text{spherical triangle on}~S_{\rm pol}^2:\nonumber\\
&~~~~~\varphi_g[C_{\rm pol}]  = \frac{1}{2}\Omega[C_{\rm pol}] \nonumber\\
&~~~~~~~~~~~~~~~~=  -\arg(\boldsymbol{E}^{(1)^\dag}\boldsymbol{E}^{(2)}
\boldsymbol{E}^{(2)^\dag} \boldsymbol{E}^{(3)}\boldsymbol{E}^{(3)^\dag}
\boldsymbol{E}^{(1)})\,.
\end{align}

\subsection{Polarisation in the eikonal limit}
The third situation we consider from the geometric phase perspective
is one that has been studied for a long time on account of its
obvious physical relevance. It is the short wave length---or eikonal
or ray---limit of Maxwell's equations, leading to differential
equations for light rays in a given transparent medium, plus the law for
evolution of the electric field along them\,\cite{born-wolf}. We first recall the
basic equations resulting from the eikonal limit, then some
important previous work, and finally consider the situation from the
geometric phase perspective.

In comparison to the previous Pancharatnam case, in the eikonal
limit the propagation (ray) direction is allowed to vary while, in a
sense to be clarified later, the polarisation state stays constant.
We consider Maxwell's equations for propagating electric (and
magnetic) fields in a transparent nonconducting non-magnetic
material medium characterised by a time-independent isotropic
refractive index function $n(\boldsymbol{x})$. To leading order, the
eikonal limit gives a system of second order ordinary differential
equations whose solutions are rays in the medium:
\begin{align}\label{eq2.27}
\frac{d}{ ds}\,\left(n(\boldsymbol{x})\frac{d\boldsymbol{x}}{ ds}\right) =
\boldsymbol{\triangledown}\,n(\boldsymbol{x})\,.
\end{align}
Each solution $\boldsymbol{x}(s)$ (for given initial conditions)
determines a ray $\Gamma$, a curve in physical three-dimensional
Euclidean space. Here $s$ is arc length measured along $\Gamma$ from
some starting point on $\Gamma$. {\em We hereafter work with some definite $\Gamma$}.

As a space curve, $\Gamma$ is characterized by the following vectors
and scalars defined pointwise along  it, the dot denoting derivative
with respect to $s$:
\begin{align}\label{eq2.28}
\boldsymbol{v}(s) &=\dot{\boldsymbol{x}}(s) = \text{unit tangent};\nonumber\\
\boldsymbol{n}(s) &=
\dot{\boldsymbol{v}}(s)/|\dot{\boldsymbol{v}}(s)|=\text{unit principal
normal};\nonumber\\
\boldsymbol{b}(s) &= \boldsymbol{v}(s)_{\wedge}\boldsymbol{n}(s) =
\text{unit binormal};\nonumber\\
\kappa(s) &=|\dot{\boldsymbol{v}}(s)| =
\text{curvature},\nonumber\\
\tau(s)&=\boldsymbol{b}(s)\cdot\dot{\boldsymbol{n}}(s)=\text{torsion}\,.
\end{align}
At each $\boldsymbol{x}(s)\in\Gamma$,
$(\boldsymbol{v}(s),\boldsymbol{n}(s),\boldsymbol{b}(s))$ is a
right handed orthonormal triad, unique at generic points with nonzero
curvature; it is locally determined by $\dot{\boldsymbol{x}}(s)$ and
$\ddot{\boldsymbol{x}}(s)$. Formally these vectors obey the
`equations of motion'
\begin{align}\label{eq2.29}
\dot{\boldsymbol{v}}  &=\kappa\,\boldsymbol{b}_\wedge \boldsymbol{v} =
(\boldsymbol{v}_\wedge\dot{\boldsymbol{v}})_\wedge\boldsymbol{v}\,,\nonumber\\
\dot{\boldsymbol{n}}  &= (\kappa\,\boldsymbol{b}+ \tau\, \boldsymbol{v})_\wedge
\boldsymbol{n}\,,\nonumber\\
\dot{\boldsymbol{b}}  &=\tau\, \boldsymbol{v}_\wedge \boldsymbol{b}.
\end{align}
The first equation (which is actually trivial) means that
$\boldsymbol{v}$ obeys the minimal Fermi-Walker transport law\,\cite{fermi-walker}, 
 while $\boldsymbol{n}$ and $\boldsymbol{b}$ do not do so.

Next we consider the evolution of the electric field
$\boldsymbol{E}(\boldsymbol{x}(s))\equiv\boldsymbol{E}(s)$ along
$\Gamma$. This comes from the next to  leading order terms in the
eikonal limit of Maxwell's equations, and when expressed in terms
of the normalised electric field $\boldsymbol{\Psi}(s)$ 
we have again the Fermi-Walker
transport law along with the transversality condition:
\begin{align}\label{eq2.30}
\boldsymbol{\Psi}(s)= \boldsymbol{E}(s) / \sqrt{\boldsymbol{E}(s)^\dag
\boldsymbol{E}(s)}\,:\quad
\dot{\boldsymbol{\Psi}}(s)&=\kappa(s)\,\boldsymbol{b}(s)_\wedge\,
\boldsymbol{\Psi}(s)\,,\nonumber\\
\boldsymbol{v}(s)\cdot\boldsymbol{\Psi}(s)&=0\,.
\end{align}
That both $\boldsymbol{v}(s)$ and $\boldsymbol{\Psi}(s)$ obey the
Fermi-Walker law is consistent with the need to maintain the
transversality condition $\boldsymbol{v}(s)\cdot\boldsymbol{\Psi}(s)
=0$ along $\Gamma$.

At each $\boldsymbol{x}(s)\in\Gamma$, $\boldsymbol{n}(s)$ and
$\boldsymbol{b}(s)$ span the transverse plane perpendicular to
$\boldsymbol{v}(s)$ there. If we introduce another orthonormal basis
in this plane, $\boldsymbol{e}_a(s)$, $a=1,2$, obeying the
Fermi-Walker transport law like $\boldsymbol{\Psi}(s)$, we have the
evolution equations
\begin{align}\label{eq2.31}
\boldsymbol{e}_a(s)\cdot\boldsymbol{e}_b(s) &=
\delta_{ab},\; \boldsymbol{e}_1(s)_\wedge
\,\boldsymbol{e}_2(s)=\boldsymbol{v}(s),\;
\boldsymbol{e}_a(s)\cdot\boldsymbol{v}(s)=0\,;\nonumber\\
\dot{\boldsymbol{e}}_a(s)&=
\kappa(s)\,\boldsymbol{b}(s)_\wedge\,\boldsymbol{e}_a(s),\,\quad a=1,2\,.
\end{align}
As initial condition we take
\begin{align}\label{eq2.32}
\boldsymbol{e}_1(s_1)=\boldsymbol{n}(s_1),\quad
\boldsymbol{e}_2(s_1)=\boldsymbol{b}(s_1)
\end{align}
at some $s=s_1$. Then the pair
$(\boldsymbol{e}_1,\boldsymbol{e}_2)$ rotates steadily with
respect to the pair $(\boldsymbol{n},\boldsymbol{b})$ at a rate given
by the torsion:
\begin{align}\label{eq2.33}
\frac{d}{ ds} \begin{pmatrix}
\boldsymbol{e}_a(s)\cdot\boldsymbol{n}(s)\\
\boldsymbol{e}_a(s)\cdot\boldsymbol{b}(s)\end{pmatrix} &=
\begin{pmatrix} 0 & \tau(s)\\ -\tau(s) & 0\end{pmatrix}
\begin{pmatrix}\boldsymbol{e}_a(s)\cdot\boldsymbol{n}(s)\\
\boldsymbol{e}_a(s)\cdot\boldsymbol{b}(s)\end{pmatrix},\nonumber\\
&\hspace{3cm} a=1,2\,;\nonumber\\
\begin{pmatrix}\boldsymbol{e}_1(s)\\
\boldsymbol{e}_2(s)\end{pmatrix}& =
\begin{pmatrix} \cos\chi (s) & -\sin\chi(s)\\ \sin\chi(s) & \cos\chi(s)\end{pmatrix}
\begin{pmatrix}\boldsymbol{n}(s)\\
\boldsymbol{b}(s)\end{pmatrix},\nonumber\\
\chi(s)& = \int_{s_1}^{s} ds^\prime \tau(s^\prime)\,.
\end{align}
Now $\boldsymbol{e}_a(s)\cdot\boldsymbol{\Psi}(s)$ are constants
along $\Gamma$:
\begin{align}\label{eq2.34}
\boldsymbol{\Psi}(s) &= z_a \boldsymbol{e}_a(s)\,,\quad z_a=
\boldsymbol{e}_a(s)\cdot\boldsymbol{\Psi}(s)=\text{constant}\,,\nonumber\\
\boldsymbol{z}^\dag\boldsymbol{z} &= (z^*_1~z_2^*)
\begin{pmatrix} z_1 \\ z_2\end{pmatrix}=1\,.
\end{align}

All the three-dimensional vectors
$\boldsymbol{x},\boldsymbol{v},\boldsymbol{n},\boldsymbol{b},
\boldsymbol{E},\boldsymbol{\Psi},\boldsymbol{e}_a$
have corresponding components with respect to some fixed global
Cartesian frame in space. The representation (\ref{eq2.34})
identifies $\boldsymbol{\Psi}(s)$ at each $\boldsymbol{x}(s)\in\Gamma$
with a `vector' $\boldsymbol{z}$ in the two-dimensional complex linear space
$\mathbb{C}^2$. Using this we can represent the polarization state
at $\boldsymbol{x}(s)\in\Gamma$ by a point $\hat{\boldsymbol{n}}(\boldsymbol{z})$
on the Poincar\'e sphere $S^2_{\rm pol}$:
\begin{align}\label{eq2.35}
\boldsymbol{\Psi}(s) \rightarrow \hat{\boldsymbol{n}}(\boldsymbol{z}) =
\boldsymbol{z}^\dag\, \boldsymbol{\tau}\, \boldsymbol{z} \in S_{\rm pol}^2\,.
\end{align}
As long as no polarization gadgets are placed anywhere on $\Gamma$,
the $z_a$ are constants, so the polarization state represented by
$\hat{\boldsymbol{n}}(\boldsymbol{z})\in S_{\rm pol}^2$ is also constant: only the
propagation direction $\boldsymbol{v}(s)$ varies. This is to be
compared with the Pancharatnam situation\,: under free propagation, 
both propagation direction
$\boldsymbol{k}$ and polarization state $\hat{\boldsymbol{n}}\in S^2_{\rm pol}$ are
constant. If polarization gadgets are placed along the axis,
$\boldsymbol{k}$ (by definition) stays constant, while $\hat{\boldsymbol{n}}(z)$
moves on $S^2_{\rm pol}$.

In the present context we can say the cyclic case occurs when the
choice of a `later' point $\boldsymbol{x}(s_2)\in \Gamma$ is such
that $\boldsymbol{v}(s_2)$, $\boldsymbol{n}(s_2)$,
$\boldsymbol{b}(s_2)$ are the same as $\boldsymbol{v}(s_1)$,
$\boldsymbol{n}(s_1)$, $\boldsymbol{b}(s_1)$ respectively at the
initial point $\boldsymbol{x}(s_1)\in\Gamma$. This happens if
\begin{align}\label{eq2.36}
\text{cyclic case}\,:\; \dot{\boldsymbol{x}}(s_2) =
\dot{\boldsymbol{x}}(s_1)\,,\quad\ddot{\boldsymbol{x}}(s_2)
=\ddot{\boldsymbol{x}}(s_1)\,.
\end{align}
The behaviours of input linear and circular polarizations are then
particularly simple. The linear case corresponds to real $z_a$; then
${\boldsymbol{\Psi}}(s)$ is a vector in space with real Cartesian
components all along $\Gamma$. From Eqs.\,(\ref{eq2.32},\ref{eq2.33})
we can relate $\boldsymbol{\Psi}(s_2)$ to ${\boldsymbol{\Psi}}(s_1)$
as follows: 
\begin{align}\label{eq2.37}
\text{Linear } & \text{ polarisation:}\,:\nonumber\\
\boldsymbol{\Psi}(s_1) &=  \cos\theta~\boldsymbol{e}_1(s_1) +
\sin\theta\,\boldsymbol{e}_2(s_1) \longrightarrow \nonumber\\
\boldsymbol{\Psi}(s_2) &=  \cos\theta~\boldsymbol{e}_1(s_2) +
\sin\theta \,\boldsymbol{e}_2(s_2)  \nonumber\\
&= \cos(\theta-\chi(s_2)) \boldsymbol{e}_1(s_1) +
\sin(\theta-\chi(s_2))\boldsymbol{e}_2(s_1)\,,\nonumber\\
\chi(s_2) &=  \int_{s_1}^{s_2} ds ~\tau(s)\,.
\end{align}
The two transverse planes at $\boldsymbol{x}(s_2)$,
$\boldsymbol{x}(s_1)$ on $\Gamma$ are parallel to one
another, and $\boldsymbol{\Psi}(s_2)$ is obtained from
$\boldsymbol{\Psi}(s_1)$ by a right handed rotation by angle
$\chi(s_2)$ about $\boldsymbol{v}(s_1)$. A detailed calculation shows
that $\chi(s_2)$ has a geometrical meaning. Over the range $s_1 \le
s \le s_2$, the unit tangent $\boldsymbol{v}(s)$ to $\Gamma$
describes a {\em closed loop} $C_{\rm dir}$ on the sphere of 
directions $S^2_{\rm dir}$\,:
\begin{align}\label{eq2.38}
C_{\rm dir} = \{\boldsymbol{v}(s) \in S^2_{\rm dir} ~|~ s_1 \le s \le s_2\} \subset
S^2_{\rm dir}, \; \boldsymbol{v}(s_2)=\boldsymbol{v}(s_1)\,.
\end{align}
Then we have the result that the integrated torsion is (the negative
of) the {\em solid angle} subtended by $C_{\rm dir}$ at 
the centre of $S^2_{\rm dir}$\,:
\begin{align}\label{eq2.39}
\chi(s_2) = \int_{s_1}^{s_2} ds ~\tau(s) = -\Omega[C_{\rm dir}]\,.
\end{align}
In the cases of circular polarisations, we get phase shifts rather
than a rotation in space. In these cases, $\boldsymbol{\Psi}(s)$ is a
complex three-vector at all $\boldsymbol{x}(s)$ on $\Gamma$\,:
\begin{align}\label{eq2.40}
{\rm RCP/LCP}\,:&\nonumber\\
\boldsymbol{\Psi}(s_1) &=
\frac{1}{\sqrt{2}}(\boldsymbol{e}_1(s_1)\pm
i\boldsymbol{e}_2(s_1))\longrightarrow\nonumber\\
\boldsymbol{\Psi}(s) &= \frac{1}{\sqrt{2}}(\boldsymbol{e_1}(s) \pm
i\boldsymbol{e}_2(s))\nonumber\\
&= \frac{1}{\sqrt{2}}\,e^{\pm
i\chi(s)}(\boldsymbol{n}(s)\pm i\boldsymbol{b}(s))\,,\nonumber\\
\boldsymbol{\Psi}(s_2) &= e^{\pm i\chi(s_2)}\boldsymbol{\Psi}(s_1)=
e^{\mp i\Omega[C_{\rm dir}]}\boldsymbol{\Psi}(s_1)\,.
\end{align}

These results on the behaviours of polarization in the ray limit of
Maxwell's equations were obtained very early by Rytov and by
Vladimirskii\,\cite{rytov,vladimirskii}. 
In particular, Rytov showed that the phase difference
between RCP and LCP evolves at a rate proportional to the torsion;
while Vladimirskii showed that the spatial rotation experienced in
the cyclic case for linear polarization is essentially by the solid
angle $\Omega[C_{\rm dir}]$.

To cast the above discussion into the geometric phase format of
Section I, it is useful to write the evolution  equation
(\ref{eq2.30}) for the (normalised) electric field in a
Schr\"odinger-like form with a suitable hermitian Hamiltonian
operator. We view $\boldsymbol{\Psi}(s)$ (referred to axes fixed in
space) as a (normalised) element of $\mathcal{H}=\mathbb{C}^3$,
which is the Hilbert space in the present context, and find:
\begin{align}\label{eq2.41}
i\,\frac{d}{ ds} \boldsymbol{\Psi}(s) &= H(s)\boldsymbol{\Psi}(s)\,,\nonumber\\
~H(s)&=i \kappa(s)\left(\boldsymbol{n}(s)\boldsymbol{v}(s)^T-
\boldsymbol{v}(s)\boldsymbol{n}(s)^T\right)\,.
\end{align}
Thus $H(s)$ is a pure imaginary antisymmetric $3\times3$ matrix. The
transversality condition $\boldsymbol{v}(s)^T\boldsymbol{\Psi}(s)=0$ is
to be added as a constraint consistent with the evolution. The
definition (\ref{eq1.7}) allows us to define a geometric phase for
any $s_1$ and $s_2$, and we find that {\em due to transversality the
dynamical phase always vanishes}. Bringing in the spaces
$\mathcal{B}_5\simeq S^5$ and $\mathcal{R}_4 = CP^{2}$, the complex
two-dimensional projective space appropriate to
$\mathcal{H}=\mathbb{C}^3$, we have:
\begin{align}\label{eq2.42}
\mathcal{C} &= \{\boldsymbol{\Psi}(s)\in \mathbb{C}^3| s_1\le s
\le
s_2\}\subset\mathcal{B}_5\,,\nonumber\\
\pi[\mathcal{C}] &= C \subset R_4\,:\nonumber\\
\varphi_g[C] &=
\varphi_{\rm tot}[\mathcal{C}]-\varphi_{\rm dyn}[\mathcal{C}]\,,\nonumber\\
\varphi_{\rm tot}[\mathcal{C}] &=\arg(\boldsymbol{\Psi}(s_1)^\dag
\boldsymbol{\Psi}(s_2))\,,\nonumber\\
\varphi_{\rm dyn}[\mathcal{C}]&= \text{Im} \int_{s_1}^{s_2}
ds\,\boldsymbol{\Psi}(s)^\dag \frac{d\boldsymbol{\Psi}(s)}{ ds} \nonumber\\
&=
\text{Im}\left(-i \int_{s_1}^{s_2} ds \,\boldsymbol{\Psi}(s)^\dag H(s)
\boldsymbol{\Psi}(s)\right)=0,\,\nonumber\\
\text{i.e.},\qquad\quad \varphi_g[C] &=
\varphi_{\rm tot}[\mathcal{C}]\,.
\end{align}
Here we recognize that $\mathcal{C}$ {\em cannot be drawn freely in
$\mathcal{B}_5$ because of the transversality condition, so in this
way it is constrained by} $\Gamma$.

To illustrate the above, let us quote some particular cases in a
table, Eq.\,(2.46)\,:\\
\begin{widetext}
\noindent
\begin{equation}\label{eq2.43}
\begin{array}{|c|c|c|c|}
\hline
~~\text{Choices of } s_1,s_2~~~~ &
~~{\text{Polarisation}}~~~ & {\text{Behaviour
of }\, \boldsymbol{\Psi}(s)} & {\varphi_g[C]}\\
\hline
\text{Free} & \text{Linear} & \text{Real} & 0 ~\text{or}~ \pi\\
\Gamma~ \text{cyclic, Eq.\,(\ref{eq2.36})} & \text{RCP/LCP} &
~~\boldsymbol{\Psi}(s_2)=e^{\mp i\Omega[C_{\rm dir}]}\boldsymbol{\Psi}(s_1)~~ &
~\;\mp \,\Omega[C_{\rm dir}]~\;\\
\hline
\end{array}
\end{equation}
\end{widetext}
The distinction between $C\subset \mathcal{R}_{4}$ and
$C_{\rm dir}\subset S^2_{\rm dir}$ should be kept in mind.

\section{Combined path and polarisation geometric phases}
The brief reviews presented in the previous Section show that the
Pancharatnam situation and the ray optic limit are mutually
complementary. In the former only the polarization state changes,
while in the latter only the propagation direction changes. Now we
try to cover the (important) middle ground between them. We endeavour to build up a
physical picture, based ultimately on Maxwell's equations, with the
motivation to arrive at geometric phases in the framework of
Section I.

As in the eikonal limit, we consider light traveling through a
transparent non-magnetic stationary medium with refractive index
$n(\boldsymbol{x})$. We recall that the concept of a single ray is
not physically meaningful and cannot be realised. The eikonal limit
of Maxwell's equations leads at first to a first order partial
differential equation in three-dimensional space for the eikonal, a
function $S(\boldsymbol{x})$. A particular eikonal $S(\boldsymbol{x})$
leads to a continuous family or succession of wave fronts over each
of which $S(\boldsymbol{x})$ is constant, and which taken together
cover some region of physical space. Rays are then lines drawn in
this region, orthogonal at each point to the wave front passing
through that point. These rays are solutions to Eq.\,(\ref{eq2.27}).
Thus one eikonal $S(\boldsymbol{x})$ determines a corresponding
succession of wave fronts and in turn one family $\mathfrak{F}_S$ of
rays. There is only one wavefront, and only one ray belonging to
$\mathfrak{F}_S$, through each point in the relevant region.

It is in this sense that single rays are not directly physically
realisable. The best that we can do is to consider a narrow or well
collimated (\,i.e., nearly parallel\,)
 bundle of nearby rays with some nonzero cross-sectional
area which may vary along the bundle 
  [\,Consequently the wavefronts along the bundle, correspondingly 
limited in their spatial extent, are nearly planar\,]. Calling this a beam, at each
`point' along it we have some finite spatially limited wavefront. In
this picture we have in mind some $\Gamma$ obeying Eq.\,(\ref{eq2.27})
acting as the `backbone' of the beam. At each location
$\boldsymbol{x}(s)\in \Gamma$, we have a propagation direction
$\boldsymbol{v}(s)$, a spatially limited `plane wave' perpendicular
to $\boldsymbol{v}(s)$, and a transverse electric field
$\boldsymbol{E}(\boldsymbol{x}(s))\equiv \boldsymbol{E}(s)$. Thus we
arrive at a physical picture of a continuous succession of limited
plane waves each at a spatial location $\boldsymbol{x}(s)$, with
propagation direction $\boldsymbol{v}(s)$ and in some polarization state. Now
we can go a step further and allow the wavelength to be finite, as
long as it is much smaller than all other physically relevant
dimensions, including the linear dimensions of the limited plane
wave elements.

In this way we motivate the passage from a physically unrealisable
ray to a realisable beam by a process of `thickening' of the
former. In the sequel, the spatial locations $\boldsymbol{x}(s)$ of
successive plane wave elements of the beam will sometimes be
omitted. The parameter $s$ continues to be distance measured along
the beam from some initial point, increasing at each location in the
direction of $\boldsymbol{v}(s)$.

For a beam propagating `freely' in the medium in this way, the
evolution equation for $\boldsymbol{\Psi}(s)$ is Eq.\,(\ref{eq2.30}).
This, as we have seen, is a consequence of Maxwell's equations in the
medium, and can be put into the Schr\"odinger-like form
(\ref{eq2.41}) with a hermitian Hamiltonian operator. The solution
Eq.\,(\ref{eq2.34}) with constant $\boldsymbol{z}$ implies a constant
polarization state $\hat{\boldsymbol{n}}(z)\in S^2_{\rm pol}$ 
given in Eq.\,(\ref{eq2.35}).

We can now go another step further and imagine placing various
polarisation gadgets over (short) stretches of the beam,
equivalently of $\Gamma$, where $\boldsymbol{z}$ varies as function
of $s$, governed by a `polarisation Hamiltonian' as in
Eq.\,(\ref{eq2.20}). Thus we arrive at new evolution equations for
$\boldsymbol{\Psi}(s)$ based on the following ingredients:
\begin{align}\label{eq3.1}
\boldsymbol{\Psi}(s) &= z_a(s)
\boldsymbol{e}_a(s)\,:\nonumber\\
&i\frac{d}{ ds}\boldsymbol{e}_a(s) =
H^{({\rm dir})}(s)\boldsymbol{e}_a(s),\nonumber\\
& H^{({\rm dir})}(s) =
i\kappa(s)\left(\boldsymbol{n}(s)\boldsymbol{v}(s)^T-
\boldsymbol{v}(s)\boldsymbol{n}(s)^T\right)\,;\nonumber\\
&i\frac{d}{ ds}\boldsymbol{z}(s) = H^{({\rm pol})}(s)\boldsymbol{z}(s),\nonumber\\
&H^{({\rm pol})}(s) =\frac {1}{2} \,\boldsymbol{\tau}\cdot
\boldsymbol{a}(s)\,,~~~\boldsymbol{a}(s) ~\text{real}\,.
\end{align}
We have now written $H^{({\rm dir})}(s)$ for the `direction' part of the
Hamiltonian, appearing in Eq.\,(\ref{eq2.41}); it is completely
determined by the local geometrical properties of $\Gamma$. The
other contribution to the evolution of $\boldsymbol{\Psi}(s)$ is from
the `polarization' part of the Hamiltonian, as in Eq.\,(\ref{eq2.20}),
written now as $H^{({\rm pol})}(s)$. This controls the evolution of the
local two-component transverse description of $\boldsymbol{\Psi}(s)$
resolved along $\boldsymbol{e}_a(s)$. The complete evolution
equation for $\boldsymbol{\Psi}(s)$ is easily found to be
Schr\"odinger-like, with a Hamiltonian which is a (complex) hermitian
$3\times3$ matrix:
\begin{align}\label{eq3.2}
i\frac{d}{ ds}\boldsymbol{\Psi}(s) &=
\left(H^{({\rm dir})}(s)+
H^{\prime\,(\rm pol)}(s)\right)\,\boldsymbol{\Psi}(s),\,\nonumber\\
H^{({\rm dir})}(s) &=
i\kappa(s)\,\left(\boldsymbol{n}(s)\boldsymbol{v}(s)^T-\boldsymbol{v}(s)
\boldsymbol{n}(s)^T\right),\,\nonumber\\
H^{\prime\,(\rm pol)}(s) &=\frac{1}{2}\,
a_j(s)(\tau_j)_{ab}\boldsymbol{e}_a(s) \boldsymbol{e}_b(s)^T.
\end{align}
The implied evolution equation for $\hat{\boldsymbol{n}}(\boldsymbol{z})
\in S^2_{\rm pol}$
is as in Eq.\,(\ref{eq2.21}):
\begin{align}\label{eq3.3}
\hat{\boldsymbol{n}}(\boldsymbol{z}(s))\equiv \hat{\boldsymbol{n}}(s)~:
\qquad \frac{d\hat{\boldsymbol{n}}(s)}{ ds} =
\boldsymbol{a}(s)_\wedge \hat{\boldsymbol{n}}(s)\,.
\end{align}

Over portions of the beam free of polarization gadgets, where
$\boldsymbol{a}(s)=0$, the local properties of $\Gamma$ determine the
propagation, and the polarization state is constant. Passage through
gadgets leads to changing $\boldsymbol{z}(s)$ and $\hat{\boldsymbol{n}}(s)$. Both
kinds of changes in $\boldsymbol{\Psi}(s)$ are ultimately traced
back to Maxwells' equations; 
 and the complete evolution equation $(\ref{eq3.2})$ respects 
the transversality condition
$\boldsymbol{v}(s)^T\,\boldsymbol{\Psi}(s)=0$. In all of this, the
separation of effects due to change in beam direction and those due
to presence of polarisation gadgets, is essentially unambiguous.

Let us now bring in geometric phase considerations. As in the ray case
in Section II(C), we are able to use the basic quantum mechanical
$\mathcal{H}-\mathcal{B}-\mathcal{R}$ framework with
$\mathcal{H}=\mathbb{C}^3$, $\mathcal{B}_5=S^5$, and
$\mathcal{R}_4=CP^2$ which is of real dimension four. For the
calculation of dynamical phases we need the result
\begin{align}\label{eq3.4}
& \text{Im} \left(\boldsymbol{\Psi}(s),\,
   \frac{d\boldsymbol{\Psi}(s)}{ ds} \right)\nonumber\\
&~~~~~= \text{Im} \text{\Large (} -i(\boldsymbol{\Psi}(s),
(H^{({\rm dir})}(s)+H^{\prime\,(\rm pol)}(s))\boldsymbol{\Psi}(s)\text{\Large )}
\text{\Large )}\nonumber\\
&~~~~~= -\frac{1}{2}\,\boldsymbol{a}(s)\cdot\hat{\boldsymbol{n}}(s)\,,
\end{align}
so {\em there is a contribution only from the presence of polarisation
gadgets} [This was to be expected since we have arranged 
the `evolution in direction' to be of vanishing dynamical phase]. 
For general $s_1$ and $s_2$ with initial and final spatial
positions $\boldsymbol{x}(s_1)$, $\boldsymbol{x}(s_2)$ on the beam we
define:
\begin{align}\label{eq3.5}
\mathcal{C} &=\{\boldsymbol{\Psi}(s) \in \mathcal{H}\,|\, s_1 \le s \le
s_2\} \subset \mathcal{B}_5\,,\nonumber\\
\pi[\mathcal{C}] &= C \subset \mathcal{R}_4\,.
\end{align}
(It is implicit that $\boldsymbol{\Psi}(s)$ is located in space at
$\boldsymbol{x}(s)$ and is transverse, so as in Section II it cannot
be drawn arbitrarily in $\mathcal{B}_5$). Then we have:
\begin{align}\label{eq3.6}
\varphi_g[{C}] &= \varphi_{\rm tot}[\mathcal{C}]-
\varphi_{\rm dyn}[\mathcal{C}]\,,\nonumber\\
\varphi_{\rm tot}[\mathcal{C}] &= \arg(\boldsymbol{\Psi}(s_1)^\dag
\boldsymbol{\Psi}(s_2))\,,\nonumber\\
\varphi_{\rm dyn}[\mathcal{C}] &=
\text{Im}\int_{s_1}^{s_2}ds~\left(\boldsymbol{\Psi}(s),\, \frac{d}{ ds}
\boldsymbol{\Psi}(s)\right)\nonumber\\
&=-\frac{1}{2} \int_{s_1}^{s_2}ds \,
\boldsymbol{a}(s)\cdot\hat{\boldsymbol{n}}(s)\,.
\end{align}

We illustrate this result in a special situation, where a connection
to the results in Section II in the Pancharatnam case (B) can be made.
Let us firstly choose $s_1$ and $s_2$ so that this stretch of
$\Gamma$ is `cyclic' in the sense of Eq.\,(\ref{eq2.36}). Then we make
an independent additional assumption that the polarisation
gadgets placed along the beam between $\boldsymbol{x}(s_1)$ and
$\boldsymbol{x}(s_2)$ are such that (\,for a particular initial
$\boldsymbol{\Psi}(s_1)\,)\,\boldsymbol{z}(s_2)$ turns out to be a phase
times $\boldsymbol{z}(s_1)$. This then means that the curve traced by
$\hat{\boldsymbol{n}}(s)\in S_{\rm pol}^2$ is a closed loop. In all the conditions
assumed are:
\begin{align}\label{eq3.7}
\dot{\boldsymbol{x}}(s_2) &=
\dot{\boldsymbol{x}}(s_1)\,,\quad\ddot{\boldsymbol{x}}(s_2) =
\ddot{\boldsymbol{x}}(s_1)\,;\nonumber\\
\boldsymbol{z}(s_2) &=e^{i\theta}\boldsymbol{z}(s_1)\,,\quad
\hat{\boldsymbol{n}}(s_2)=\hat{\boldsymbol{n}}(s_1)\,;\nonumber\\
C_{\rm pol} &=\{\hat{\boldsymbol{n}}(s) \in S^2_{\rm pol}|s_1\le s \le s_2\} \subset
S^2_{\rm pol},\;\,\text{closed}\,.
\end{align}
By Eqs.\,(\ref{eq2.33},\ref{eq2.39}) we relate
$\boldsymbol{e}_a(s_2)$ to $\boldsymbol{e}_a(s_1)$:
\begin{align}\label{eq3.8}
\begin{pmatrix}
\boldsymbol{e}_1(s_2)\\
\boldsymbol{e}_2(s_2)\end{pmatrix} = \begin{pmatrix}
\cos\Omega[C_{\rm dir}] & \sin\Omega[C_{\rm dir}]\\
-\sin\Omega[C_{\rm dir}] & \cos\Omega[C_{\rm dir}]\end{pmatrix} \begin{pmatrix}
\boldsymbol{e}_1(s_1)\\ \boldsymbol{e}_2(s_1)\end{pmatrix}\,,
\end{align}
where $\Omega[C_{\rm dir}]$ is the solid angle subtended at the centre of
$S_{\rm dir}^2$ by $C_{\rm dir}$ defined in Eq.\,(\ref{eq2.38}). We see that with the
conditions (\ref{eq3.7}) we deal with {\em two closed loops}, $C_{\rm dir}\subset
S^2_{\rm dir}$ and $C_{\rm pol}\subset S^2_{\rm pol}$, 
on the sphere of directions and on
the Poincar\'e sphere respectively. Now we can calculate the
geometric phase for this situation using Eq.\,(\ref{eq3.6}):
\begin{align}\label{eq3.9}
\varphi_g[C] &= \arg\left(\boldsymbol{z}(s_1)^\dag \begin{pmatrix}
\cos\Omega[C_{\rm dir}] & -\sin\Omega[C_{\rm dir}]\\ \sin\Omega[C_{\rm dir}] &
\cos\Omega[C_{\rm dir}]\end{pmatrix}
e^{i\theta}\boldsymbol{z}(s_1)\right)\nonumber\\
&~~~~+ \frac{1}{2}\int_{s_1}^{s_2} ds\,
\boldsymbol{a}(s)\cdot\hat{\boldsymbol{n}}(s)\nonumber\\
&= \theta+\frac{1}{2}\int_{s_1}^{s_2} ds\,\boldsymbol{a}(s)\cdot
\hat{\boldsymbol{n}}(s)\nonumber\\
&~~~~+ \arg\text{\Large (}\cos\Omega[C_{\rm dir}]+2i\sin\Omega[C_{\rm dir}]~\text{Im}
z_1(s_1)z_2(s_1)^*\text{\Large )}.
\end{align}
On comparing Eqs.\,(\ref{eq2.24},\ref{eq2.25}) of the Pancharatnam
situation with Eq.\,(\ref{eq3.1}), we see that the first two terms
here add up to $\frac{1}{2}\Omega[C_{\rm pol}]$:
\begin{align}\label{eq3.10}
\theta +\frac{1}{2} \int_{s_1}^{s_2} ds \, \boldsymbol{a}(s)
\cdot\hat{\boldsymbol{n}}(s)
= \frac{1}{2}\Omega[C_{\rm pol}]\,.
\end{align}
If we finally specialize to input circular polarisations, the
third term also simplifies:
\begin{align}\label{eq3.11}
{\rm RCP/LCP}\,:\quad z_1(s_1) &= \frac{1}{\sqrt{2}}\,,\quad
z_2(s_1) = \pm \frac{i}{\sqrt{2}}\,;\nonumber\\
\arg \text{\Large (}\cos\Omega[C_{\rm dir}] & 
+2i\sin\Omega[C_{\rm dir}]\,\text{Im}~z_1(s_1)z_2(s_1)^*\text{\Large )}\nonumber\\
&~~~~~~~~= \mp \Omega[C_{\rm dir}]\,,
\end{align}
and then the geometric phase becomes,
\begin{align}\label{eq3.12}
\varphi_g[C] = \frac{1}{2}\Omega[C_{\rm pol}]\mp\,\Omega[C_{\rm dir}]\,.
\end{align}

We may remark finally that while the geometric phase in the
present physical situation is always defined by Eq.\,(\ref{eq3.6}), it
is only in a quite special situation that we get a simple final
expression (\ref{eq3.12}), in a way combining the Pancharatnam
result (\ref{eq2.25}) and the pure ray result (\ref{eq2.43}). What
needs to be stressed however is that the separation of the
contributions from the sphere of directions $S^2_{\rm dir}$ and from the
Poincar\'e (polarization) sphere $S^2_{\rm pol}$ is essentially unambiguous.
In the Pancharatnam limit, $C_{\rm dir}$ shrinks to a point and we
recover (\ref{eq2.25}); while in the pure ray limit with no
polarisation gadgets, it is $C_{\rm pol}$ that shrinks to a point and we
get back (\ref{eq2.43}) for circular polarisations.

\section{Some global aspects of the sphere of directions}
The situation analysed in the previous Section from the geometric
phase point of view is that of a (narrow well collimated) light beam
of fixed frequency travelling in physical space through a given
transparent medium, encountering various polarisation gadgets on its
way. The path of the beam is based on a ray
$\Gamma=\{\boldsymbol{x}(s)\}\subset\mathcal{R}^3$ obeying
Eq.\,(\ref{eq2.27}) for a given refractive index function
$n(\boldsymbol{x})$. From $\Gamma$ we obtain a particular
one-dimensional curve $C_{\rm dir}=\{\,\boldsymbol{v}(s)=\dot{\boldsymbol{x}}(s)\,\}
\subset S^2_{\rm dir}$, the two-dimensional sphere of directions.

The ray $\Gamma$ also gives a preferred choice of a real orthonormal
basis $\{\boldsymbol{e}_a(s)\}$ in the transverse plane at each
$\boldsymbol{x}(s)\in\Gamma$, perpendicular to $\boldsymbol{v}(s)$
there. By resolving the normalised complex transverse electric field
$\boldsymbol{\Psi}(\boldsymbol{x}(s))$ with respect to this basis, we
are able to describe it by a normalized complex two-component column
vector $\boldsymbol{z}(s)$, leading to the representation of the
polarization state by a point $\hat{\boldsymbol{n}}(s)\in 
S^2_{\rm pol}$. In particular,
real $\boldsymbol{\Psi}(\boldsymbol{x}(s))$ implies real
$\boldsymbol{z}(s)$ and vice versa, corresponding to linear
polarisations.

The choice of $\Gamma$ thus provides both $C_{\rm dir} \subset S_{\rm dir}^2$, and
$\{\boldsymbol{e}_a(s)\}$. We can regard the latter as a preferred
real orthonormal basis in the real tangent plane
$T_{\boldsymbol{v}(s)}S^2_{\rm dir} \simeq \mathbb{R}^2$, for each
$\boldsymbol{v}(s)\in C_{\rm dir}$. As a result, the geometric phase
contributions from beam direction and beam polarisation are
essentially unambiguously separated.

Let us now view the problem from another {\em more global perspective,
not immediately related to a ray or to a picture embedded in 
physical space}. We take the sphere $S^2_{\rm dir}$ of plane wave propagation
directions as starting point, writing $\hat{\boldsymbol{k}}$ for points on it
(instead of $\boldsymbol{v}(s)$ obtained from $\Gamma$ as 
upto now). Each $\hat{\boldsymbol{k}}$ is
the unit vector in the direction of a wave vector $\boldsymbol{k}$
associated with a possible (spatially limited) propagating plane
wave. We now ask if there is a way to choose a real orthonormal
basis $\{\boldsymbol{e}_a(\hat{\boldsymbol{k}})\}$ in the real tangent plane
$T_{\hat{\boldsymbol{k}}}S^2_{\rm dir}\simeq\mathbb{R}^2$, well defined and varying
smoothly with $\hat{\boldsymbol{k}}$ for all $\hat{\boldsymbol{k}} 
\in S^2_{\rm dir}$.

Since this question is posed prior to the possible choice of a ray
$\Gamma$, even if such $\{\boldsymbol{e}_a(\hat{\boldsymbol{k}})\}$ exist, it
need have nothing to do with the $\{\boldsymbol{e}_a(s)\}$ later
supplied by a ray $\Gamma$ at a point on it where
$\boldsymbol{v}(s)=\hat{\boldsymbol{k}}$. As we have seen, it is
$\{\boldsymbol{e}_a(s)\}$ which has specific advantages from a
physical point of view, which may be absent with
$\{\boldsymbol{e}_a(\hat{\boldsymbol{k}})\}$.

It is however a known  fact from differential geometry that such
choices of $\{\boldsymbol{e}_a(\hat{\boldsymbol{k}})\}$ 
for all $\hat{\boldsymbol{k}} \in
S^2_{\rm dir}$ do not exist. This is expressed by saying that the sphere
$S^2_{\rm dir}$ is not parallelizable\,\cite{combing}\,---as a 
real four-dimensional manifold
the tangent bundle $TS^2_{\rm dir}$ is not (homeomorphic to) the product
$S^2_{\rm dir}\times\mathbb{R}^2$. A useful way to display this
circumstance, suited for further developments, is as follows.

In real three-dimensional Euclidean space let us choose a right
handed Cartesian system of axes with origin $O$, and with
$\hat{\boldsymbol{e}}_j$, $j=1,2,3$, the unit vectors along the coordinate axes.
Points on the unit sphere $S^2_{\rm dir}$ with centre at $O$ will be written
$\hat{\boldsymbol{k}} =(\sin\theta\cos\phi, \sin\theta\sin\phi,\cos\theta)$,
$0\le\theta\le\pi$, $0\le\phi<2\pi$. It is necessary to define two
subsets $S^2_N$, $S^2_S$ of $S^2_{\rm dir}$ whose union gives $S^2_{\rm dir}$ but
which have a nontrivial (indeed, substantial) overlap: 
\begin{align}\label{eq4.0}
&S^2_N =\{\hat{\boldsymbol{k}}(\theta,\phi)\in S^2_{\rm dir}\,|\, 
0\le \theta <\pi\,,\quad
0\le\phi <2\pi\}\,,\nonumber\\
&S^2_S =\{\hat{\boldsymbol{k}}(\theta,\phi)\in S^2_{\rm dir}
\,|\, 0<  \theta \le\pi\,,\quad
0\le\phi <2\pi\}\,,\nonumber\\
&S^2_N \cup S^2_S = S^2_{\rm dir}\,,\nonumber\\
&S^2_N \cap S^2_S 
 = \{\hat{\boldsymbol{k}}(\theta,\phi)\in S^2_{\rm dir}\,|\, 0 < \theta <
\pi\,,\, 0\le \phi <2\pi\}\,.
\end{align}
We need the action of proper rotations, elements of the rotation group $SO(3)$, 
on $S_{\rm dir}^2$. The right handed rotation about 
axis $\hat{\boldsymbol{a}}\in S^2$ by angle
$\alpha$ corresponds to the $3\times 3$ matrix
\begin{align}\label{eq4.1}
\mathcal{R}_{jk}(\hat{\boldsymbol{a}},\alpha)& =\delta_{jk} \cos\alpha +
a_ja_k(1-\cos\alpha) - \epsilon_{jkl}a_l\sin\alpha\,,\nonumber\\
 &~~~~~~~~~~~~~~~~~~~~~~~~~0\le\alpha \le 2\pi\,.
\end{align}
For any $\hat{\boldsymbol{k}}\in S_{\rm dir}^2$, there are infinitely many rotations
carrying $\hat{\boldsymbol{e}}_3$ to $\hat{\boldsymbol{k}}$. 
However there is no way to choose
one such rotation for each $\hat{\boldsymbol{k}}$, such that it is globally
well-defined and varies smoothly with $\hat{\boldsymbol{k}}$ for 
all $\hat{\boldsymbol{k}} \in
S^2_{\rm dir}$. Over $S^2_N$, which is $S^2_{\rm dir}$ with just one 
point (the south pole) removed, a convenient choice does exist\,:
\begin{align}\label{eq4.3}
A^\prime(\hat{\boldsymbol{k}}) &=
\mathcal{R}(\hat{\boldsymbol{e}}_3,\phi)\mathcal{R}(\hat{\boldsymbol{e}}_2,\theta)
\mathcal{R}(\hat{\boldsymbol{e}}_3,\phi)^{-1}\nonumber\\
&= \mathcal{R}(\hat{\boldsymbol{e}}_2\cos\phi-
\hat{\boldsymbol{e}}_1\sin\phi,\theta)\,,\nonumber\\
&~~~~~0\le\theta<\pi,\;0\le\phi<2\pi\,;\nonumber\\
A^\prime(\hat{\boldsymbol{k}})\hat{\boldsymbol{e}}_3 &= \hat{\boldsymbol{k}}\,.
\end{align}
This is well defined at $\theta=0$ but not at $\theta=\pi$. Acting
on $\hat{\boldsymbol{e}}_1,\hat{\boldsymbol{e}}_2$ at the North pole, we get a real
orthonormal basis for the tangent plane $T_{\hat{\boldsymbol{k}}}S^2_{\rm dir}$ when
$\theta<\pi$:
\begin{align}\label{eq4.4}
\boldsymbol{e}_1^\prime(\hat{\boldsymbol{k}}) 
=  A^\prime(\hat{\boldsymbol{k}})\hat{\boldsymbol{e}}_1 &=
\begin{pmatrix}\sin^2\phi+\cos\theta\cos^2\phi\\
(\cos\theta-1)\sin\phi\cos\phi\\
-\sin\theta\cos\phi\end{pmatrix}\,,\nonumber\\
\boldsymbol{e}_2^\prime(\hat{\boldsymbol{k}}) 
= A^\prime(\hat{\boldsymbol{k}})\hat{\boldsymbol{e}}_2 &=
\begin{pmatrix} (\cos\theta-1)\sin\phi\cos\phi\\
\cos^2\phi+\cos\theta\sin^2\phi\\ -\sin\theta\sin\phi\end{pmatrix}
,\;\hat{\boldsymbol{k}}\in S^2_N\,.
\end{align}
Over $S^2_S$ a similar choice is:
\begin{align}\label{eq4.5}
A^{\prime\prime}(\hat{\boldsymbol{k}}) &=
R(\hat{\boldsymbol{e}}_3,\phi){R}(\hat{\boldsymbol{e}}_2,\theta)
{R}(\hat{\boldsymbol{e}}_3,\phi)\,,\nonumber\\
&~~~~~~~~~~~0<\theta\le\pi\,,\quad 0 \le\phi<2\pi\,:\nonumber\\
&~~~~~A^{\prime\prime}(\hat{\boldsymbol{k}})\hat{\boldsymbol{e}}_3 
= \hat{\boldsymbol{k}}\,.
\end{align}
Now this is well defined at $\theta=\pi$ but not at $\theta=0$.
Acting on $\hat{\boldsymbol{e}}_1,\hat{\boldsymbol{e}}_2$ at the 
North pole we get a different
real orthonormal basis for $T_{\hat{\boldsymbol{k}}}S^2_{\rm dir}$ when $\theta>0$:
\begin{align}\label{eq4.6}
\boldsymbol{e}_1^{\prime\prime}(\hat{\boldsymbol{k}}) =
A^{\prime\prime}(\hat{\boldsymbol{k}})\hat{\boldsymbol{e}}_1 &=
\begin{pmatrix}\cos\theta\cos^2\phi-\sin^2\phi\\
(1+\cos\theta)\sin\phi\cos\phi\\
-\sin\theta\cos\phi\end{pmatrix}\,,\nonumber\\
\boldsymbol{e}_2^{\prime\prime}(\hat{\boldsymbol{k}}) =
A^{\prime\prime}(\hat{\boldsymbol{k}})\hat{\boldsymbol{e}}_2 &=
\begin{pmatrix} -(1+\cos\theta)\sin\phi\cos\phi\\
\cos^2\phi-\cos\theta\sin^2\phi\\ \sin\theta\sin\phi\end{pmatrix}
,\;\hat{\boldsymbol{k}}\in S^2_S\,.
\end{align}
In the overlap, which is all of $S^2_{\rm dir}$ with just the north 
and south poles removed, we have connecting or `transition' formulae:
\begin{align}\label{eq4.7}
\hat{\boldsymbol{k}}\in S^2_N \cap S^2_S \,:& \nonumber\\
  A^{\prime\prime}(\hat{\boldsymbol{k}}) &=
A^\prime(\hat{\boldsymbol{k}})R(\hat{\boldsymbol{e}}_3,2\phi) =
R(\hat{\boldsymbol{k}},2\phi)A^\prime(\hat{\boldsymbol{k}})\,;\nonumber\\
\boldsymbol{e}_a^{\prime\prime}(\hat{\boldsymbol{k}})&
 = R(\hat{\boldsymbol{k}},2\phi)\boldsymbol{e}_a^{\prime}
(\hat{\boldsymbol{k}}),~~a=1,2.
\end{align}

There are now two equally good ways  to express the nonparallelizable
nature of $S^2_{\rm dir}:$ (i) it is not possible to extend the definition
of $A^\prime(\hat{\boldsymbol{k}})$ (respectively $A^{\prime\prime}
(\hat{\boldsymbol{k}})$) to cover
the South pole $\theta=\pi$ (respectively the North pole $\theta=0$)
possessing smooth behaviour for all 
$\hat{\boldsymbol{k}}\in S^2_{\rm dir}$; (ii) the real
orthonormal bases $\{\boldsymbol{e}_a^\prime(\hat{\boldsymbol{k}})\}$,
$\{\boldsymbol{e}_a^{\prime\prime}(\hat{\boldsymbol{k}})\}$ for
$T_{\hat{\boldsymbol{k}}}S^2_{\rm dir}$ over $S^2_N$, $S^2_S$ respectively cannot be
modified in any way to yield a real orthonormal basis for
$T_{\hat{\boldsymbol{k}}}S^2_{\rm dir}$ varying smoothly 
with $\hat{\boldsymbol{k}}$ all over $S^2_{\rm dir}$.
A more formal statement is this: it is impossible to find two
smoothly varying angles 
$\chi^{\,'}(\hat{\boldsymbol{k}}),\,\chi^{\,''}(\hat{\boldsymbol{k}})$ 
over $S^2_N$, $S^2_S$ respectively 
such that the transition group element $R(\hat{\boldsymbol{e}}_3,2\phi)$ in
Eq.\,(\ref{eq4.7}) can be factorised as
\begin{align}\label{eq4.8}
R(\hat{\boldsymbol{e}}_3,2\phi) = R(\hat{\boldsymbol{e}}_3,
\chi^\prime(\hat{\boldsymbol{k}}))
R(\hat{\boldsymbol{e}}_3,\chi^{\prime\prime}(\hat{\boldsymbol{k}}))^{-1}, 
~\forall\hat{\boldsymbol{k}}\in
S^2_N\cap S^2_S.
\end{align}
For, if such choices were possible, then $A^\prime(\hat{\boldsymbol{k}})$
$R(\hat{\boldsymbol{e}}_3$, $\chi^\prime(\hat{\boldsymbol{k}}))
=A^{\prime\prime}(\hat{\boldsymbol{k}})$
$R(\hat{\boldsymbol{e}}_3,\chi^{\prime\prime}(\hat{\boldsymbol{k}}))$ 
would carry $\hat{\boldsymbol{e}}_3$
to $\hat{\boldsymbol{k}}$ and be smoothly defined 
for all $\hat{\boldsymbol{k}} \in S^2_{\rm dir}$.

Now the `topological obstruction' described above is in the {\em real
domain}, i.e., viewing each tangent plane 
$T_{\hat{\boldsymbol{k}}}S^2_{\rm dir}$ as a
real two-dimensional vector space $\mathbb{R}^2$. It has however
been pointed out recently that if one complexifies each
$T_{\hat{\boldsymbol{k}}}S^2_{\rm dir}$ into a complex two-dimensional vector space
$(T_{\hat{\boldsymbol{k}}}S^2_{\rm dir})^c\simeq \mathbb{C}^2$, then the obstruction
vanishes\,\cite{rnss}\,: it is possible to choose orthonormal bases for these
complexified tangent spaces in a globally smooth manner. There is
naturally considerable freedom in such choices; we describe now a
group theory based choice which seems natural and minimal in some
sense. This requires the use of the group $SU(2)$ (which is a double
cover of $SO(3)$, though this property is not used in 
the $SU(2)$ version of Eq.\,(4.8) established below). 
What we will show is that {\em the factorisation
attempted in Eq.\,(\ref{eq4.8}) is possible if on the right hand side
we allow for elements from $SU(2)$}.

The defining representation of $SU(2)$ is
\begin{align}\label{eq4.9}
&SU(2) = \nonumber\\
&\{\mathcal{U}=2\times 2 \text{ complex matrices}~|~\mathcal{U}^\dag\mathcal{U}
  = 1\!\!1_{2 \times 2},\text{det }\mathcal{U}=1\},
\end{align}
group composition being matrix multiplication. The axis-angle
description of $SU(2)$ elements is
\begin{align}\label{eq4.10}
\mathcal{U}(\hat{\boldsymbol{a}},\alpha)& 
=e^{-\frac{i}{2}\alpha\hat{\boldsymbol{a}}\cdot\boldsymbol{\sigma}}
= \cos\frac{\alpha}{2}\,1\!\!1 - i\hat{\boldsymbol{a}}\cdot\boldsymbol{\sigma}
\sin\frac{\alpha}{2}\,,\nonumber\\
&~~~~~~~~~~~~~ \hat{\boldsymbol{a}}\in S^2\,,\, 0\le \alpha\le 2\pi\,.
\end{align}
 The two-to-one mapping $SU(2)\to SO(3)$ respecting the composition
laws, the homomorphism, is:
\begin{align}\label{4.11}
\mathcal{U}(\hat{\boldsymbol{a}},\alpha) \in SU(2) \longrightarrow  
R(\hat{\boldsymbol{a}},\alpha)\in
SO(3)\,.
\end{align}
The rotations $A^\prime(\hat{\boldsymbol{k}})$, 
$A^{\prime\prime}(\hat{\boldsymbol{k}})$
defined in Eqs.\,(\ref{eq4.3},\,\ref{eq4.5}) are images, in the sense of
this mapping, of elements $\mathcal{U}^\prime(\hat{\boldsymbol{k}})$,
$\mathcal{U}^{\prime\prime}(\hat{\boldsymbol{k}})$ in $SU(2)$ respectively:
\begin{align}\label{eq4.12}
\mathcal{U}^\prime(\hat{\boldsymbol{k}}) &= e^{-\frac{i}{2}\phi\sigma_3}
e^{-\frac{i}{2}\theta\sigma_2} e^{\frac{i}{2}\phi\sigma_3},~~\hat{\boldsymbol{k}}\in
S_N^2\,;\nonumber\\
\mathcal{U}^{\prime\prime}(\hat{\boldsymbol{k}}) &= e^{-\frac{i}{2}\phi\sigma_3}
e^{-\frac{i}{2}\theta\sigma_2} e^{-\frac{i}{2}\phi\sigma_3},~~\hat{\boldsymbol{k}}\in
S_S^2\,.
\end{align}
Now the overlap transition rule (\ref{eq4.7}) involves the subgroup
of elements $R(\hat{\boldsymbol{e}}_3,2\phi)\in SO(2) \subset SO(3)$, which happen to
`coincide' with elements $\mathcal{U}(\hat{\boldsymbol{e}}_2,4\phi)\in SU(2)$ in
the following sense:
\begin{align}\label{eq4.13}
R(\hat{\boldsymbol{e}}_3,2\phi) &= 
\left( \begin{array}{cc}
\mathcal{U}(\hat{\boldsymbol{e}}_2,4\phi) & \begin{array}{c} 0\\0\end{array}\\
\begin{array}{cc} ~0~&~0~ \end{array} & 1
\end{array}\right)\,,\nonumber\\
\mathcal{U}(\hat{\boldsymbol{e}}_2,4\phi) &=e^{-2i\phi\sigma_2} =
\begin{pmatrix} 
\cos2\phi & -\sin2\phi\\ 
\sin2\phi &\cos2\phi
\end{pmatrix}\,\,.
\end{align}
It now turns out that \textit{within} $SU(2)$  a factorisation of
the form (\ref{eq4.8}) {\em is possible}:
\begin{align}\label{eq4.14}
\mathcal{U}(\hat{\boldsymbol{e}}_2, 4\phi) &=
{\cal V}^\prime(\hat{\boldsymbol{k}})^{-1}\,
{\cal V}^{\prime\prime}(\hat{\boldsymbol{k}})\,,\nonumber\\
{\cal V}^\prime(\hat{\boldsymbol{k}}) &= e^{-i\phi\sigma_2}
e^{-i \frac{\theta}{2}\sigma_1}
e^{i\phi\sigma_2}\,,\quad \hat{\boldsymbol{k}}\in S^2_N\,;\nonumber\\
{\cal V}^{\prime\prime}(\hat{\boldsymbol{k}}) &=e^{-i\phi\sigma_2}
e^{-i\frac{\theta}{2}\sigma_1} e^{-i\phi\sigma_2}\,,\quad \hat{\boldsymbol{k}}\in
S^2_S\,.
\end{align}

The structures of ${\cal V}^\prime(\hat{\boldsymbol{k}})$, 
${\cal V}^{\prime\prime}(\hat{\boldsymbol{k}})$
are suggested by those of ${\cal U}^{\,'}(\hat{\boldsymbol{k}})$, 
${\cal U}^{\,''}(\hat{\boldsymbol{k}})$ in
Eq.\,(\ref{eq4.12}): in the latter we make the cyclic changes
$\sigma_1\to\sigma_3\to\sigma_2\to\sigma_1$, and replace $\phi$ by
$2\phi$. If we use Eq.\,(\ref{eq4.14}) in Eq.\,(\ref{eq4.13}) and then
in Eq.\,(\ref{eq4.7}) we see that
\begin{align}\label{eq4.15}
A^{\prime\prime}(\hat{\boldsymbol{k}}) &= A^\prime(\hat{\boldsymbol{k}}) 
\left( \begin{array}{cc}
{\cal V}^\prime(\hat{\boldsymbol{k}})^{-1}  
{\cal V}^{\prime\prime}(\hat{\boldsymbol{k}}) & \begin{array}{c} 0\\0\end{array}\\
\begin{array}{cc} ~0~~&~~0~ \end{array} & 1
\end{array}\right)\,,\nonumber\\
\text{i.e.,}~\mathcal{A}(\hat{\boldsymbol{k}}) & = A^\prime(\hat{\boldsymbol{k}})
\left( \begin{array}{cc}
{\cal V}^\prime(\hat{\boldsymbol{k}})^{-1}  & \begin{array}{c} 0\\0\end{array}\\
\begin{array}{cc} ~0~&~0~ \end{array} & 1
\end{array}\right)\nonumber\\
&= A^{\prime\prime}(\hat{\boldsymbol{k}})
\left( \begin{array}{cc}
 {\cal V}^{\prime\prime}(\hat{\boldsymbol{k}})^{-1} & 
\begin{array}{c} 0\\0\end{array}\\
\begin{array}{cc} ~0~~&~~0~ \end{array} & 1
\end{array}\right)
\end{align}
is a globally well-defined and smoothly varying matrix in $SU(3)$
with the property
\begin{align}\label{eq4.16}
\mathcal{A}(\hat{\boldsymbol{k}})\hat{\boldsymbol{e}}_3 = 
\hat{\boldsymbol{k}}\,, ~~\forall~~\hat{\boldsymbol{k}}\in S^2_{\rm dir}\,\,.
\end{align}
Here, the group $SU(3)$ is the three-dimensional extension of
$SU(2)$ in Eq.\,(\ref{eq4.9}), and consists of $3\times 3$ unitary
unimodular matrices. The subset {\em (not subgroup)} of $SU(3)$ carrying
$\hat{\boldsymbol{e}}_3$ to $\hat{\boldsymbol{k}}$ is easy to characterise\,:
\begin{align}\label{eq4.17}
\mathcal{A}\in SU(3)\,:\;
\mathcal{A}\hat{\boldsymbol{e}}_3&=\hat{\boldsymbol{k}}\Leftrightarrow
\mathcal{A}=A
\left( \begin{array}{cc}
{\cal U} & \begin{array}{c} 0\\0\end{array}\\
\begin{array}{cc} ~0~~&~~0~ \end{array} & 1
\end{array}\right)\,,\nonumber\\
&~~A\in SO(3),\; \mathcal{U}\in SU(2),\;
A\hat{\boldsymbol{e}}_3=\hat{\boldsymbol{k}}\,.
\end{align}
(This decomposition is however not unique on account of the shared
elements (\ref{eq4.13})). And indeed $\mathcal{A}(\hat{\boldsymbol{k}})$ is of
this form, and becomes after simplification:
\begin{align}\label{eq4.18}
\mathcal{A}(\hat{\boldsymbol{k}})=R(\hat{\boldsymbol{e}}_3,
\phi)R(\hat{\boldsymbol{e}}_2,\theta)
\left( \begin{array}{cc}
e^{\frac{i}{2}\theta\sigma_1}  e^{i\phi\sigma_2} & \begin{array}{c} 0\\0\end{array}\\
\begin{array}{cc} ~0~~&~~0~ \end{array} & 1
\end{array}\right).
\end{align}

If we act with $\mathcal{A}(\hat{\boldsymbol{k}})$ on 
$\hat{\boldsymbol{e}}_1$, $\hat{\boldsymbol{e}}_2$ at
the North pole we obtain a globally well defined and smooth complex
orthonormal basis for $(T_{\hat{\boldsymbol{k}}} S^2_{\rm dir})^c$ 
all over $S_{\rm dir}^2$. To distinguish these 
vectors from the real ones encountered up to
now we write them as $\boldsymbol{g}_a(\hat{\boldsymbol{k}})$, $a=1,2$.
They are simply related to $\boldsymbol{e}_a^\prime(\hat{\boldsymbol{k}})$,
$\boldsymbol{e}_a^{\prime\prime}(\hat{\boldsymbol{k}})$ of Eqs.\,(\ref{eq4.4},
\ref{eq4.6}):
\begin{align}\label{eq4.19}
\boldsymbol{g}_1(\hat{\boldsymbol{k}}) 
&=\mathcal{A}(\hat{\boldsymbol{k}})\hat{\boldsymbol{e}}_1 =
\cos\theta/2~ \boldsymbol{e}^\prime_1(\hat{\boldsymbol{k}})+i\sin\theta/2\,
\boldsymbol{e}_2^{\prime\prime}(\hat{\boldsymbol{k}})\,,\nonumber\\
\boldsymbol{g}_2(\hat{\boldsymbol{k}}) 
&=\mathcal{A}(\hat{\boldsymbol{k}})\hat{\boldsymbol{e}}_2 =
\cos\theta/2~ \boldsymbol{e}^\prime_2(\hat{\boldsymbol{k}})+i\sin\theta/2\,
\boldsymbol{e}_1^{\prime\prime}(\hat{\boldsymbol{k}})\,;\nonumber\\
\hat{\boldsymbol{k}}&\cdot \boldsymbol{g}_a(\hat{\boldsymbol{k}})=0\,,\quad
\boldsymbol{g}_a(\hat{\boldsymbol{k}})^*\cdot\boldsymbol{g}_b(\hat{\boldsymbol{k}})
=\delta_{ab}\,.
\end{align}
While it should be evident {\em a priori} that, given the existence of such
global complex $\boldsymbol{g}_a(\hat{\boldsymbol{k}})$ 
all over $S^2_{\rm dir}$, there
should be considerable freedom in their choice, the specific form of
$\mathcal{A}(\hat{\boldsymbol{k}})$ in Eq.\,(\ref{eq4.18}) suggests that the choice
(\ref{eq4.19}) is specially simple. 
 In particular, for the one-parameter family of 
$\hat{\boldsymbol{k}}(\theta,\phi)$ with fixed $\phi$ the connection 
between $\boldsymbol{g}_a(\hat{\boldsymbol{k}})$ 
and $(\boldsymbol{e}_a^{\prime}(\hat{\boldsymbol{k}}),
\,\boldsymbol{e}_a^{\prime\prime}(\hat{\boldsymbol{k}}))$ is 
through (portion of) a {\em one-parameter subgroup}  
of $SU(2)$. Hereafter we always use
$\{\boldsymbol{g}_a(\hat{\boldsymbol{k}})\}$ given above.

Any (complex) three-vector $\boldsymbol{\psi}(\hat{\boldsymbol{k}})$ orthogonal to
$\hat{\boldsymbol{k}}$, thus belonging 
to $(T_{\hat{\boldsymbol{k}}}S^2_{\rm dir})^c$, can be expanded
as
\begin{align}\label{eq4.20}
\boldsymbol{\psi}(\hat{\boldsymbol{k}}) 
= z_a \boldsymbol{g}_a(\hat{\boldsymbol{k}}), \; z_a
= \boldsymbol{g}_a(\hat{\boldsymbol{k}})^*\cdot\boldsymbol{\psi}
    (\hat{\boldsymbol{k}}),\;
\boldsymbol{z} = \begin{pmatrix}z_1\\ z_2\end{pmatrix}
\in\mathbb{C}^2\,.
\end{align}
With respect to $\{\boldsymbol{g}_a(\hat{\boldsymbol{k}})\}$, and \textit{as a
convention}, we can regard $\boldsymbol{\psi}(\hat{\boldsymbol{k}})$ 
at $\hat{\boldsymbol{k}}$
and $\boldsymbol{\psi}^\prime(\hat{\boldsymbol{k}}^\prime)$ 
at $\hat{\boldsymbol{k}}^\prime$ as
`the same' if
\begin{align}\label{eq4.21}
\boldsymbol{g}_a(\hat{\boldsymbol{k}})^*
\cdot\boldsymbol{\psi}(\hat{\boldsymbol{k}}) = 
\boldsymbol{g}_a(\hat{\boldsymbol{k}}^\prime)^* 
\cdot \boldsymbol{\psi}^{\,'}(\hat{\boldsymbol{k}}^{\,'}) = z_a\,.
\end{align}
In this sense we see that the union of 
$(T_{\hat{\boldsymbol{k}}}S^2_{\rm dir})^c$ over
all $\hat{\boldsymbol{k}}$ is a Cartesian product, which as discussed earlier is
not true for the usual tangent bundle in the real domain:
\begin{align}\label{eq4.22}
TS^2_{\rm dir} &= 
\bigcup_{\hat{\boldsymbol{k}}\in S^2_{\rm dir}} 
T_{\hat{\boldsymbol{k}}}S^2_{\rm dir} \not\simeq
S^2_{\rm dir} \times\mathbb{R}^2\,,\nonumber\\
(TS^2_{\rm dir})^c &\equiv \bigcup_{\hat{\boldsymbol{k}}
\in S^2_{\rm dir}} (T_{\hat{\boldsymbol{k}}}S^2_{\rm dir})^c
\simeq S^2_{\rm dir} \times \mathbb{C}^2\,.
\end{align}
The most obvious use of the above result is in the following context. Suppose
$\boldsymbol{\psi}(\hat{\boldsymbol{k}})$ is a complex vector-valued transverse
function of $\hat{\boldsymbol{k}}$, which for concreteness we regard as an
element of a Hilbert space $\mathcal{H}$ as follows:
\begin{align}\label{eq4.23}
\mathcal{H}&=\{\,\boldsymbol{\psi}(\hat{\boldsymbol{k}})\in\mathbb{C}^3 \,\,|\,\, 
\hat{\boldsymbol{k}}\in
S^2_{\rm dir},\; \hat{\boldsymbol{k}}\cdot\boldsymbol{\psi}
(\hat{\boldsymbol{k}})=0,\;\nonumber\\
&~~~\|\boldsymbol{\psi}\|^2 = \int d\Omega(\hat{\boldsymbol{k}})
\boldsymbol{\psi}(\hat{\boldsymbol{k}})^* 
\cdot\boldsymbol{\psi}(\hat{\boldsymbol{k}}) <
\infty\,\}\,,
\end{align}
with $d\Omega(\hat{\boldsymbol{k}})=\sin\theta d\theta d\phi$ the solid angle over
$S^2_{\rm dir}$. Then we can expand $\boldsymbol{\psi}(\hat{\boldsymbol{k}})$ 
in the basis
(\ref{eq4.19}) and have:
\begin{align}\label{eq4.24}
\boldsymbol{\psi}(\hat{\boldsymbol{k}}) &=
z_a(\hat{\boldsymbol{k}})\boldsymbol{g}_a(\hat{\boldsymbol{k}})\,,\quad
z_a(\hat{\boldsymbol{k}})=\boldsymbol{g}_a
(\hat{\boldsymbol{k}})^*\cdot\boldsymbol{\psi}(\hat{\boldsymbol{k}})\,,\nonumber\\
\|\boldsymbol{\psi}\|^2 &= \int d\Omega (\hat{\boldsymbol{k}})
\boldsymbol{z}(\hat{\boldsymbol{k}})^\dag \boldsymbol{z}(\hat{\boldsymbol{k}})\,.
\end{align}
This shows that $\mathcal{H}$ is the tensor product
\begin{align}\label{eq4.25}
\mathcal{H} = L^2(S^2_{\rm dir})\otimes\mathcal{H}^{(2)}\,\,,
\end{align}
where $L^2(S^2_{\rm dir})$ is the Hilbert space of (scalar) complex square
integrable functions over $S^2_{\rm dir}$, and $\mathcal{H}^{(2)}$ is the
two-dimensional complex Hilbert space (appropriate for the `polarization qubit').

If $\boldsymbol{E}(\hat{\boldsymbol{k}})$ is a transverse electric field amplitude
of a plane wave with propagation direction $\hat{\boldsymbol{k}}$, using the
expansion (\ref{eq4.20}) we may attempt to represent its
polarisation state by a point on the Poincar\'e sphere $S^2_{\rm pol}$ in
the `usual' way:
\begin{align}\label{eq4.26}
\hat{\boldsymbol{k}}\cdot\boldsymbol{E}(\hat{\boldsymbol{k}})=0:
\;\; &\boldsymbol{E}(\hat{\boldsymbol{k}})
= z_a\boldsymbol{g}_a(\hat{\boldsymbol{k}}),\;
z_a=\boldsymbol{g}_a(\hat{\boldsymbol{k}})^*\cdot\boldsymbol{E}(\hat{\boldsymbol{k}})
\nonumber\\
&~~\rightarrow \hat{\boldsymbol{n}}(\boldsymbol{z}) 
= (\boldsymbol{z}^\dag\boldsymbol{z})^{-1}\boldsymbol{z}^\dag
\boldsymbol{\tau}\boldsymbol{z} \in
S^2_{\rm pol}\,.
\end{align}
However this is not in general the `usual representation' of
polarisation states in the sense that, for instance, linear
polarisations corresponding to real $\boldsymbol{E}(\hat{\boldsymbol{k}})$ (upto
overall phases) need not imply real $\boldsymbol{z}$, so
$\hat{\boldsymbol{n}}(\boldsymbol{z})$ may not lie on the 
equator of $S^2_{\rm pol}$ in the
1-2 plane. Indeed, for $\hat{\boldsymbol{k}}\in S_N^2$, from
Eqs.\,(\ref{eq4.7},\ref{eq4.19}) we have:
\begin{align}\label{eq4.27}
\begin{pmatrix} \boldsymbol{e}^\prime_1(\hat{\boldsymbol{k}})\\
\boldsymbol{e}^\prime_2(\hat{\boldsymbol{k}})\end{pmatrix} = 
{\cal U}_0(\theta,\,\phi) \,
\begin{pmatrix}\boldsymbol{g}_1(\hat{\boldsymbol{k}})\\
\boldsymbol{g}_2(\hat{\boldsymbol{k}})\end{pmatrix},&\nonumber\\
{\cal U}_0(\theta,\,\phi) = \begin{pmatrix}
\cos\frac{\theta}{2} + i\sin\frac{\theta}{2}\sin2\phi &
-i\sin\frac{\theta}{2}\cos2\phi\\ -i\sin\frac{\theta}{2}\cos2\phi &
\cos\frac{\theta}{2}-i\sin\frac{\theta}{2}\sin2\phi\end{pmatrix};&\nonumber\\
\frac{1}{\sqrt{2}}(\boldsymbol{e}_1^\prime(\hat{\boldsymbol{k}})
+i\boldsymbol{e}^\prime_2(\hat{\boldsymbol{k}}))=
\frac{1}{\sqrt{2}}\left(\cos\frac{\theta}{2}+e^{2i\phi}\sin\frac{\theta}{2}\right)
\boldsymbol{g}_1(\hat{\boldsymbol{k}}) &~~~\nonumber\\
+ \frac{i}{\sqrt{2}}\left(\cos\frac{\theta}{2}
- e^{2i\phi}
\sin\frac{\theta}{2}\right)\boldsymbol{g}_2(\hat{\boldsymbol{k}});&\nonumber\\
\frac{1}{\sqrt{2}}(\boldsymbol{e}_1^\prime(\hat{\boldsymbol{k}})
-i\boldsymbol{e}^\prime_2(\hat{\boldsymbol{k}}))=
 \frac{1}{\sqrt{2}}\left(\cos\frac{\theta}{2}-e^{-2i\phi}\sin\frac{\theta}{2}\right)
\boldsymbol{g}_1(\hat{\boldsymbol{k}}) &~~~\nonumber\\
- \frac{i}{\sqrt{2}}\left(\cos\frac{\theta}{2}
+ e^{-2i\phi} \sin\frac{\theta}{2}\right)\boldsymbol{g}_2(\hat{\boldsymbol{k}}).&
\end{align}
Using these in Eq.\,(\ref{eq4.26}) we find that states of RCP and LCP
are represented on $S^2_{\rm pol}$ by the diametrically opposite points
$\pm(\sin\theta\cos2\phi$, $\sin\theta\sin2\phi,\cos\theta)$, not by
the usual North and South poles $(0,0,\pm1)$. Correspondingly linear
polarization states lie on the great circle on $S^2_{\rm pol}$ 
 in the plane orthogonal to
$(\sin\theta\cos2\phi, \sin\theta \sin2\phi, \cos\theta)$.

\section{Global bases and geometric phases}
We consider applications of the global results of the previous
Section to the calculation of geometric phases. In the treatment in
Sections II, III the starting point was a ray $\Gamma$ in a given
transparent medium, based on which a beam passing through
polarisation gadgets was then considered. From this, a curve
$C_{\rm dir}\subset S^2_{\rm dir}$ was obtained, as in Eq.\,(\ref{eq2.38}). The
normalised electric field along the beam was then used to define a
curve $\mathcal{C}\subset \mathcal{B}_5$  in the quantum mechanical
$\mathcal{H}-\mathcal{B}-\mathcal{R}$ framework, Eqs.\,(\ref{eq2.42},
\ref{eq3.5}), with $\mathcal{H}=\mathbb{C}^3$. At all stages the
validity of Maxwell's equations was kept in mind.

In \cite{rnss}, however,  a curve $C_{\rm dir}\subset S^2_{\rm dir}$ is taken as the
starting point for the discussion of geometric phases for beams with
varying direction and polarisation state. From the point of view
developed by us, this would mean that in principle, for a chosen
$C_{\rm dir}\subset S^2_{\rm dir}$ to be physically realisable, we must imagine a
transparent medium with suitable refractive index function $n(\boldsymbol{x})$,
and a ray $\Gamma$ in this medium, such that a beam traveling along
$\Gamma$ reproduces $C_{\rm dir}$ as we follow
$\boldsymbol{v}(s)=\dot{\boldsymbol{x}}(s)$ along $\Gamma$. All this
as well as the validity of Maxwell's equations will be implicitly
assumed in what follows.

In the notation of Section IV, then, we imagine being given a curve
$C_{\rm dir}=\{\hat{\boldsymbol{k}}(s)
\in S^2_{\rm dir}\}\subset S^2_{\rm dir}$, and at each value of $s$
a normalised transverse electric field $\boldsymbol{\Psi}(s)$:
\begin{align}\label{eq5.1}
\boldsymbol{\Psi}(s) \in \mathbb{C}^3,\;
\boldsymbol{\Psi}(s)^*\cdot \boldsymbol{\Psi}(s) = 1, \;
\hat{\boldsymbol{k}}(s)\cdot\boldsymbol{\Psi}(s)=0\,.
\end{align}
(Though not explicitly stated, the parameter $s$ could be the
distance measured from some starting point on a beam in physical
space $\mathbb{R}^3$). For calculating geometric phases we again use
the $\mathcal{H}-\mathcal{B}-\mathcal{R}$ framework with
$\mathcal{H}=\mathbb{C}^3$, (the framework used in \cite{rnss} is
different and is briefly recounted in the Appendix), and define
\begin{align}\label{eq5.2}
\mathcal{C} = \{\boldsymbol{\Psi}(s)\in
\mathcal{H}\,&|\,\boldsymbol{\Psi}(s)^\dag\boldsymbol{\Psi}(s)=1,\nonumber\\
&~~~\hat{\boldsymbol{k}} \cdot \boldsymbol{\Psi}(s)=0\,,
\quad s_1\le s\le s_2\}\subset
\mathcal{B}_5\,,\nonumber\\
\pi[\mathcal{C}] &= C \subset\mathcal{R}_4\,.
\end{align}
We expand $\boldsymbol{\Psi}(s)$ in the complex global basis for
$(T_{\hat{\boldsymbol{k}}}S^2_{\rm dir})^c$ described in Eq.\,(\ref{eq4.19}):
\begin{align}\label{eq5.3}
\boldsymbol{\Psi}(s) &= z_a(s)\boldsymbol{g}_a(\hat{\boldsymbol{k}}(s)),\;
z_a(s)=\boldsymbol{g}_a(\hat{\boldsymbol{k}}(s))^*
  \cdot\boldsymbol{\Psi}(s),\,\nonumber\\
\boldsymbol{z}(s) &= \begin{pmatrix}z_1(s)\\
z_2(s)\end{pmatrix},\; \boldsymbol{z}(s)^\dag \boldsymbol{z}(s) =
1\,.
\end{align}
To compute the dynamical phase $\varphi_{\rm dyn}[\mathcal{C}]$ we need
as ingredients:
\begin{align}\label{eq5.4}
\dot{\boldsymbol{\Psi}}(s) &= \frac{d}{ds}\,\boldsymbol{\Psi}(s) = 
\dot{{z}}_a(s) \,\boldsymbol{g}_a(\hat{\boldsymbol{k}}(s)) 
+ {z}_a(s)\, \dot{\boldsymbol{g}}_a(\hat{\boldsymbol{k}}(s)),\nonumber\\
(\boldsymbol{\Psi}(s),\dot{\boldsymbol{\Psi}}(s)) &=
\boldsymbol{z}(s)^\dag\dot{\boldsymbol{z}}(s) + z_a(s)^*
\boldsymbol{g}_a(\hat{\boldsymbol{k}}(s))^*
\cdot \dot{\boldsymbol{g}}_b(\hat{\boldsymbol{k}}(s))
z_b(s).
\end{align}
The second term leads us to define a $2\times 2$ hermitian matrix
$h(s)$ as
\begin{align}\label{eq5.5}
h_{ab}(s) = h_{ba}(s)^* = - i\boldsymbol{g}_a(\hat{\boldsymbol{k}}(s))^*\cdot
\dot{\boldsymbol{g}}_b(\hat{\boldsymbol{k}}(s))\,,
\end{align}
and then we have:
\begin{align}\label{eq5.6}
\varphi_g[C] &= \varphi_{\rm tot}[\mathcal{C}] -
\varphi_{\rm dyn}[\mathcal{C}]\,,\nonumber\\
\varphi_{\rm tot}[\mathcal{C}] &= \arg(\boldsymbol{\Psi}(s_1)^*\cdot
\boldsymbol{\Psi}(s_2))\,,\nonumber\\
\varphi_{\rm dyn}[\mathcal{C}] &= \text{Im} \int_{s_1}^{s_2}
ds~(\boldsymbol{\Psi}(s)\,,\dot{\boldsymbol{\Psi}}(s))\,\nonumber\\
&= \text{Im}\left\{\int_{s_1}^{s_2} ds\,
\boldsymbol{z}(s)^\dag\dot{\boldsymbol{z}}(s) + i\int_{s_1}^{s_2} ds
\, \boldsymbol{z}(s)^\dag h(s)\boldsymbol{z}(s)\right\}\,.
\end{align}
With some algebra the elements of $h(s)$ can be calculated in terms
of $\theta(s)$, $\phi(s)$, the spherical polar angles of
$\hat{\boldsymbol{k}}(s)$, and their derivatives $\dot{\theta}(s)$,
$\dot{\phi}(s)$:
\begin{align}\label{eq5.7}
h_{11}(s) &= -h_{22}(s) \nonumber\\
&= -\frac{1}{2}\left[ \dot{\theta}(s)\,\sin2\phi(s)
+ \dot{\phi}(s)\,\cos2\phi(s) \,\sin2\theta(s)\right],\nonumber\\
h_{12}(s) &= h_{21}(s)^* \nonumber\\
&=\frac{1}{2}\left[\dot{\theta}(s)\,\cos2\phi(s) -
( \sin2\theta(s)\,\sin2\phi(s)\right.\nonumber\\
&~~~~~~~~ ~~~~~~~~ \left. + {i}(\,1-\cos2\theta(s)\,)\dot{\phi}(s)\,\right].
\end{align}
It is interesting that the elements of the matrix $h(s)$, which
arise from the dependences of $\boldsymbol{g}_a(\hat{\boldsymbol{k}})$ on
$\hat{\boldsymbol{k}}$, have rather elementary forms, which can be ascribed to
the group theoretical arguments that led to the construction of
$\{\boldsymbol{g}_a(\hat{\boldsymbol{k}})\}$.

As an illustration, let us consider the case where $C_{\rm dir}$ is a closed
loop, i.e., $\hat{\boldsymbol{k}}(s_2)=\hat{\boldsymbol{k}}(s_1)$. 
Let us further assume that
$\boldsymbol{\Psi}(s_2)$ differs from $\boldsymbol{\Psi}(s_1)$ just by
a phase $\theta$ so that ${\cal C}$ is closed. Since in any case 
$\boldsymbol{g}_a(\hat{\boldsymbol{k}})$'s are 
determined by $\hat{\boldsymbol{k}}$, these
assumptions mean that
\begin{align}\label{eq5.8}
\boldsymbol{g}_a(\hat{\boldsymbol{k}}(s_2)) &=
\boldsymbol{g}_a(\hat{\boldsymbol{k}}(s_1))\,;\nonumber\\
\boldsymbol{\Psi}(s_2) &= e^{i\theta} \boldsymbol{\Psi}(s_1)
\Rightarrow \,\boldsymbol{z}(s_2) = e^{i\theta}
\boldsymbol{z}(s_1),\; \hat{\boldsymbol{n}}(s_2) = \hat{\boldsymbol{n}}(s_1).
\end{align}
Thus $\hat{\boldsymbol{n}}(s)\equiv \hat{\boldsymbol{n}}(\boldsymbol{z}(s))$ 
describes a closed
loop $C_{\rm pol}\subset S^2_{\rm pol}$, and the geometric phase (\ref{eq5.6})
becomes:
\begin{align}\label{eq5.9}
\varphi_g[C] &= \theta - {\rm Im}\int_{s_1}^{s_2}
ds\,\boldsymbol{z}(s)^\dag\dot{\boldsymbol{z}}(s) - \int_{s_1}^{s_2}
ds\, \boldsymbol{z}(a)^\dag h(s)\boldsymbol{z}(s).
\end{align}
Comparing the first two terms with Eq.\,(\ref{eq2.25}) we see that
they reproduce exactly $\frac{1}{2}\Omega[C_{\rm pol}]$, and the net result is
\begin{align}\label{eq5.10}
\varphi_g[C] = \frac{1}{2}\Omega[C_{\rm pol}] - \int_{s_1}^{s_2} ds \,
\boldsymbol{z}(s)^\dag h(s)\boldsymbol{z}(s)\,.
\end{align}
Further simplification of the second term seems not possible on
general grounds, unless one has some information on the way
$\boldsymbol{\Psi}(s)$ varies with $s$ as $\hat{\boldsymbol{k}}(s)$ traces the
loop $C_{\rm dir}$.

The separation of $\varphi_g[C]$ into the two terms on the right in
Eq.\,(\ref{eq5.10}) corresponds to the use of the
$\{\boldsymbol{g}_a(\hat{\boldsymbol{k}})\}$ as a basis 
for $(T_{\hat{\boldsymbol{k}}}S^2_{\rm dir})^c$
at each $\hat{\boldsymbol{k}}$. A change from 
$\{\boldsymbol{g}_a(\hat{\boldsymbol{k}})\}$ to
some other globally smooth basis would alter both terms, while
preserving the value of $\varphi_g[C]$. This could possibly limit the direct
physical meaning we may ascribe to, say, $\frac{1}{2}\,\Omega[C_{\rm pol}]$ on
the right hand side.

\section{Concluding Remarks}
We hope to have shown that in all geometric phase considerations 
in the domain of classical optics, the mathematical framework of quantum 
mechanics is adequate and flexible enough to provide a basis 
for the entire analysis. This is so in scalar wave, pure polarisation, as well as 
beam propagation problems. We have attempted to provide a clear physical 
picture of the situations being considered, fully tracing the phenomena 
ultimately to  Maxwell's equations in every case. 
The relevance of global topological aspects when discussing 
propagation direction and polarisation state 
simultaneously was pointed out in \cite{rnss}.  
 In our treatment we have addressed these using elementary group theoretical 
arguments relevant to the situation\,---\,leading, in our view, 
to particularly simple and elegant results. 

The approach of this work now needs to be extended to other situations 
where, in place of a narrow beam endowed with polarisation properties, 
an extended polarised wave field in space is contemplated. This and other 
similar extensions will be taken up elsewhere. 
    





\section*{Appendix\,:~Comparison with the approach in \cite{rnss}}
\renewcommand{\theequation}{A.\,\arabic{equation}}

\setcounter{equation}{0}


Throughout this paper we have tried to show that the standard
$\mathcal{H}-\mathcal{B}-\mathcal{R}$ structure of quantum mechanics,
with $\mathcal{B}$ a $U(1)$ principal fibre bundle over base
$\mathcal{R}$, can be used under all circumstances to handle
geometric phases in classical optical situations. In \cite{rnss} a
somewhat different structure has been used. We describe here briefly
the connection between the two approaches.

For geometric phases associated with light beams we have used the
complex three-dimensional Hilbert space
$\mathcal{H}\simeq\mathbb{C}^3$ with inner product; the unit
sphere $\mathcal{B}_5\simeq S^5$ of real dimension five; and the ray
space $\mathcal{R}_4\simeq CP^2$ of real dimension four. Here
$\mathcal{B}_5$ is a $U(1)$ principal fibre bundle over base
$\mathcal{R}_4$. We now define characteristic subsets of these
spaces as follows:
\begin{align}\label{a.1}
\hat{\boldsymbol{k}}\in S^2_{\rm dir}&:\nonumber\\
\mathcal{H}_{\hat{\boldsymbol{k}}} 
&= \{\boldsymbol{E} \in \mathcal{H} \,|\, \hat{\boldsymbol{k}}
\cdot\boldsymbol{E}= 0 \} \simeq \mathbb{C}^2\,;\nonumber\\
\mathcal{B}_{\hat{\boldsymbol{k}}} 
&= \mathcal{B}_5 \cap\mathcal{H}_{\hat{\boldsymbol{k}}} =
\{\boldsymbol{E}\in \mathcal{H} \,|\,
\boldsymbol{E}^\dag\boldsymbol{E}=1\,,\quad
\hat{\boldsymbol{k}}\cdot\boldsymbol{E}=0\}\,;\nonumber\\
\mathcal{R}_{\hat{\boldsymbol{k}}} &= \mathcal{B}_{\hat{\boldsymbol{k}}}/U(1) =
 \{\rho(\boldsymbol{E}) = 
\boldsymbol{E}\boldsymbol{E}^\dag 
\in \mathcal{R}_4\,| \,\boldsymbol{E}
\in \mathcal{B}_{\hat{\boldsymbol{k}}}\}\,;\nonumber\\
 \mathcal{B}_{\hat{\boldsymbol{k}}} &\simeq  S^3 \,,~~
 \mathcal{R}_{\hat{\boldsymbol{k}}}\simeq S^2,~~\text{for each } 
\hat{\boldsymbol{k}} \in S^2_{\rm dir}\,.
 \end{align}
In more detail in the case of $\mathcal{R}_{\hat{\boldsymbol{k}}}$ we have:
\begin{align}\label{a.2}
\rho \in \mathcal{R}_{\hat{\boldsymbol{k}}} &\Leftrightarrow \,\rho =3\times 3
~\text{complex matrix,}\nonumber\\
&~~~~~\rho^\dag =
\rho^2=\rho\ge0\,,~ {\rm Tr}\,\rho = 1\,,~\rho\hat{\boldsymbol{k}}=0\,.
\end{align}
For two points $\hat{\boldsymbol{k}}$, $\hat{\boldsymbol{k}}^\prime
\in S^2_{\rm dir}$, we find easily:
\begin{align}\label{a.3}
\hat{\boldsymbol{k}}_\wedge \hat{\boldsymbol{k}}^\prime \ne 0: 
\mathcal{B}_{\hat{\boldsymbol{k}}}\cap
\mathcal{B}_{\hat{\boldsymbol{k}}^\prime} = \{\boldsymbol{E} = q
\frac{\hat{\boldsymbol{k}}_\wedge\hat{\boldsymbol{k}}^\prime}{|
\hat{\boldsymbol{k}}_\wedge\hat{\boldsymbol{k}}^\prime|}
\,,~|q|=1\}\,,
\end{align}
consisting of essentially real $\boldsymbol{E}$ corresponding to
linear polarisations.

In contradistinction, the  total and base spaces 
used in \cite{rnss} are $\mathcal{T},
\mathcal{L}$ defined as:
\begin{align}\label{a.4}
\mathcal{T} = \bigcup_{\hat{\boldsymbol{k}}\in
S^2_{\rm dir}}~\mathcal{B}_{\hat{\boldsymbol{k}}}~;~~~~\text{(a)}\nonumber\\
\mathcal{L} = \bigcup_{\hat{\boldsymbol{k}}\in
S^2_{\rm dir}}~\mathcal{R}_{\hat{\boldsymbol{k}}}~\cdot~~~~\text{(b)}
\end{align}
These  too are of real dimensions five and four respectively, and
$\mathcal{T}$ is a $U(1)$ principal fibre bundle over base
$\mathcal{L}$\,. 

It is easy to see that the second statement in Eq.\,\eqref{eq4.22} 
leads to related Cartesian product structures for ${\cal T}$ and ${\cal L}$\,:
\begin{align}\label{a.4b}
{\cal T} = S^{2}_{\rm dir} \times S^3,~~{\cal L} 
= S^{2}_{\rm dir} \times S^{2}_{\rm pol}.
\end{align}
In our treatment, as mentioned above, we use uniformly 
${\cal B}_5$ (the sphere of normalised vectors in $\mathbb{C}^3$) rather
than $\mathcal{T}$, and the associated projective space 
$\mathcal{R}_4 \equiv CP^2$ rather than $\mathcal{L}$.
It is important to recognize that $\mathcal{T}\ne\mathcal{B}_5$, and
$\mathcal{L}\ne \mathcal{R}_4$. Writing $p$ for general points in
$\mathcal{T}$:
\begin{align}\label{a.5}
p \in \mathcal{T} \Leftrightarrow ~p=(\hat{\boldsymbol{k}},
\boldsymbol{E})\,,\qquad \hat{\boldsymbol{k}}\in S^2_{\rm dir}\,,
\quad \boldsymbol{E}\in
\mathcal{B}_{\hat{\boldsymbol{k}}}\,.
\end{align}
Since $\mathcal{B}_{\hat{\boldsymbol{k}}}\subset \mathcal{B}_5$, the map
$\mathcal{T}\to \mathcal{B}_5$ is well-defined:
\begin{align}\label{a.6}
p = (\hat{\boldsymbol{k}}, \boldsymbol{E}) \in \mathcal{T} \rightarrow
\boldsymbol{E}\in \mathcal{B}_5\,.
\end{align}
However this is a {\em many-to-one map}. Given
$\boldsymbol{E}\in\mathcal{B}_5$, $p=(\hat{\boldsymbol{k}},\boldsymbol{E})$ is not
unique as:
\begin{align}\label{a.7}
\text{if}~\boldsymbol{E}_\wedge\boldsymbol{E}^* \ne 0~ :~ &
\hat{\boldsymbol{k}}~\text{is fixed upto a sign},\nonumber\\ 
&~~~\text{resulting in a two-fold ambiguity};\nonumber\\
\text{if}~\boldsymbol{E}_\wedge\boldsymbol{E}^* = 0~:~ & \hat{\boldsymbol{k}}~
\text{is fixed upto an $SO(2)$ rotation},\nonumber\\ 
&~~~\text{more precisely an $O(2)$ rotation},\nonumber\\
 &~~~\text{resulting in a continuous ambiguity}\nonumber\\ 
 &~~~ \text{involving linear polarisation states.}
\end{align}
Thus, while both $\mathcal{T}$ and $\mathcal{B}_5$ are 
real five-dimensional manifolds, we do not have a one-to-one map between them,
so they are not identical spaces. In a similar way, it can be
checked that $\mathcal{L}$ and $\mathcal{R}_4$ are nonidentical.

In \cite{rnss}, geometric phases are defined for smooth closed curves
$\mathcal{C}_0\subset\mathcal{T}$, with images $C_0\subset\mathcal{L}$.
Such a curve $\mathcal{C}_0$ in parametrised form may be written as
\begin{align}\label{a.8}
\mathcal{C}_0 = \left\{ p(s) = \left( \hat{\boldsymbol{k}}(s), \,
\boldsymbol{E}(s)\right) \in {\cal T} \, |\, s_1 \leq s \leq s_2 \right\} 
\subset {\cal T}\,,
\end{align}
with suitable end point conditions. In our approach, 
since as seen in Eq.\,\eqref{a.6} the map ${\cal T} \to {\cal B}_5$ is 
well-defined, we can pass from $\mathcal{C}_0 \subset {\cal T}$ to $\mathcal{C} 
\subset {\cal B}_5$ in an unambiguous manner\,:
\begin{align}\label{a.9}
\mathcal{C} = \left\{ \boldsymbol{E}(s) \in {\cal B}_{\hat{\boldsymbol{k}}(s)} 
\, |\, s_1 \leq s \leq s_2 \right\} \subset {\cal B}_5\,,
\end{align}
and then use Eq.\,\eqref{eq1.7} to define the geometric phase in the 
kinematic approach. 
 This is similar to the way in which in
Sections 2 and 3 we take the electric field vector along a ray or a beam and use
it to obtain a smooth curve in ${\cal B}_5$ for which a geometric phase
can be defined using the kinematic approach. 
The expression for the   phase given in\,\cite{rnss} is  
the same as 
in our treatment, which stays entirely within  the standard 
$\mathcal{H}-\mathcal{B}-\mathcal{R}$ structure of quantum mechanics.


\begin{thebibliography}{}
\bibitem{inner-product}
Inner products in Hilbert spaces will be generally written as $(\psi,\,\phi)$ 
rather than as $\langle\psi|\phi\rangle$ in Dirac notation. 
For two-component complex column vectors 
$\boldsymbol{z},\,\boldsymbol{z}^\prime$ 
we sometimes write 
$\boldsymbol{z}^{\prime\, \dagger} \boldsymbol{z}$ in place of 
$(\boldsymbol{z}^\prime,\, \boldsymbol{z})$. For three-component 
complex column vectors, described by Cartesian components in 
physical space for instance, we write  
$(\boldsymbol{\psi},\,  \boldsymbol{\phi})$ or 
$\boldsymbol{\psi}^\dagger \boldsymbol{\phi}$  or 
$\boldsymbol{\psi}^* \cdot \boldsymbol{\phi}$ as convenient.
\bibitem{comment-su(2)}
The definition of this group is given later, in Eq.\,(4.9).
\bibitem{comment-so(3)}
The group $SO(3)$ is defined  in the axis-angle description later in Eq.\,(4.2).
\bibitem{comment-homo}
The relationship between $SU(2)$ and $SO(3)$ is described, using axis-angle
variables, later in Eq.\,(4.11) below.
\bibitem{comment-arcs}
It may be useful to recall that for a spherical triangle on $S^2$, the 
corresponding solid angle (subtended at the centre of the sphere) is the 
`spherical excess', i.e., the amount by which the sum of the three 
internal angles exceeds $\pi$. This excess occurs because $S^2$ possesses 
positive curvature.
\end{thebibliography}
\end{document}